\pdfoutput=1
\documentclass[11pt]{article}

\usepackage{graphicx}
\usepackage{amsmath}
\usepackage{amssymb}
\usepackage{dcolumn}
\usepackage{bm}
\usepackage{slashed}
\usepackage{color}
\usepackage{authblk}
\usepackage{ulem}
\usepackage{CJK}

\def\l{{\lambda}}
\def\L{{\Lambda}}
\def\d{{\delta}}
\def\D{{\Delta}}
\def\o{{\omega}}

\def\e{{\epsilon}}

\def\a{{\alpha}}
\def\b{{\beta}}
\def\c{{\chi}}

\def\g{{\gamma}}
\def\G{{\Gamma}}

\def\h{\eta}
\def\p{{\pi}}

\def\m{{\mu}}
\def\n{{\nu}}
\def\r{{\rho}}
\def\s{{\sigma}}
\def\S{{\Sigma}}

\def\th{{\theta}}

\def\ps{{\psi}}

\def\z{{\zeta}}
\def\({\left(}
\def\){\right)}
\def\[{\left[}
\def\]{\right]}

\newcommand{\pd}{{\partial}}
\newcommand{\dg}{\dagger}

\newcommand{\tr}{\text{tr}}

\newcommand{\bzero}{{(0)}}
\newcommand{\bone}{{(1)}}
\newcommand{\qst}{q^*}

\date{\today}

\begin{document}

\begin{CJK}{UTF8}{gbsn}

\title{\bf Spin Polarized Quasi-particle in Off-equilibrium Medium}

\author[1]{{Shu Lin}
\thanks{linshu8@mail.sysu.edu.cn}}
\affil[1]{School of Physics and Astronomy, Sun Yat-Sen University, Zhuhai 519082, China}
\author[1,2]{{Jiayuan Tian}
\thanks{tian12@iu.edu}}
\affil[2]{Physics Department, Indiana University, Bloomington, Indiana 47405, USA}

\maketitle


\begin{abstract}
	It is usually believed that physics in off-equilibrium state characterized by hydrodynamic gradient can be equivalently studied using equilibrium state with suitable metric perturbation. We scrutinize this assumption using chiral kinetic theory in curved space, focusing on spin response to hydrodynamic gradient. Two effects of metric perturbation have been identified: one is to change particle motion by scattering it off the metric fields, which does capture spin response to hydrodynamic gradient, but is limited by kinematic condition in the scattering picture. The other is the genuine effect of off-equilibrium state, which is realizable by mapping the equilibrium state in curved space to flat space through a suitable frame choice. It lifts the kinematic condition in the spin response to hydrodynamic gradient. We classify off-equilibrium effect on spin polarization into modifications of (i) spectral function; (ii) distribution function; (iii) KMS relation. While the last two have been studied using chiral kinetic theory, the first one is usually ignored in kinetic description. We perform a detailed analysis on the first one, finding the radiative correction to spectral function leads to a polarized quasi-particle. The degeneracy of spin responses to different hydrodynamic sources at tree-level is also lifted by radiative correction. 
\end{abstract}

\newpage

\section{Introduction and Summary}

Spin polarization is one of the most important observables in particle physics. It offers unique probe of production mechanism and internal structure of particles. In relativistic heavy ion collisions, a novel mechanism of spin polarization due to initial orbital angular momentum in off-central collisions has been proposed \cite{Liang:2004ph}. It is expected to generate spin polarization of baryon and spin alignment of vector meson \cite{Liang:2004ph,Liang:2004xn}. An outstanding feature of spin polarization in heavy ion collision (HIC) is the role of quark-gluon plasma (QGP) medium undergoing rapid evolution, which provides a valuable probe to the spinning property of QGP. Over the past decade, a variety of spin polarization phenomena have been observed in heavy ion experiments \cite{STAR:2017ckg}. One of the best understood example among them is the global polarization of $\L$ hyperon, which characterizes the vorticity of QGP \cite{Becattini:2013fla,Fang:2016vpj,Li:2017slc,Wei:2018zfb,Liu:2019krs,Becattini:2017gcx}. The local spin polarization of $\L$ hyperon and vector meson spin alignment allow us to study spin induced by more general off-equilibrium effect described by hydrodynamic gradient \cite{Hidaka:2017auj,Liu:2021uhn,Becattini:2021suc,Fu:2021pok,Becattini:2021iol,Yi:2021ryh,Li:2022vmb,Wagner:2022gza} and fluctuations effect \cite{Sheng:2022wsy,Sheng:2022ffb,Muller:2021hpe,Gao:2021rom}. 

Understanding the massive precise data requires a theoretical framework that can treat spin degree of freedom (DOF) systematically. Spin-averaged kinetic theory has been widely used in studying transport coefficients in weakly coupled QGP. The generalization to spin DOF leads to quantum kinetic theory (QKT), see \cite{Hidaka:2022dmn} for a recent review. One advantage of kinetic theory framework is that it allows for an efficient treatment of interaction via collision term, which could be technically much harder in field theory based approach \cite{Gagnon:2006hi,Gagnon:2007qt}. Incorporation of collision term in QKT has been made recently in \cite{Yang:2020hri,Weickgenannt:2021cuo,Sheng:2021kfc,Lin:2021mvw} and phenomenological studies have been carried out in \cite{Fang:2022ttm,Wang:2022yli,Lin:2022tma,Fang:2023bbw,Fang:2024vds,Lin:2024zik}. Apart from the collisional effect, there is yet another type of correction from interaction, that is the modification of spectral function. The two types of effect contribute at different orders: effects of collision and modified spectral function are realized through two-loop and one-loop self-energies of particles \cite{Lin:2021mvw}. Despite the two-loop self-energies are formally suppressed by $g^4$, it is enhanced by a factor of $1/g^4$ \cite{Lin:2022tma,Lin:2024zik} from large deviation of steady state from equilibrium. It follows that the collision effect dominates over effect of modified spectral function, which occurs at $O(g^2)$. For this reason, most phenomenological implementations are based on free particles with leading order collision term. Radiative correction to collisions have been considered in \cite{Yamamoto:2023okm,Fang:2023bbw}, see also \cite{Ghiglieri:2018dib} for spin averaged transports.

In a broader context, we may pose the following question: what is the spin polarization in an off-equilibrium medium characterized by hydrodynamic gradient. Instead of dealing with off-equilibrium state directly, one often mimics the off-equilibrium state by perturbing equilibrium state with external sources. For off-equilibrium characterized by hydrodynamic gradient, the corresponding source is metric perturbation. This reduces the off-equilibrium field theory problem into an equilibrium field theory problem with metric perturbation. This approach has been pursued for polarization recently in \cite{Liu:2021uhn}.

While the idea seems natural, the equivalence between the two approaches remains an assumption. We will scrutinize the validity of this assumption within the QKT framework. With quasi-particles as the DOF, metric fields can change the particle motion by scattering, and simultaneously change the definition of correlation function. In this work, we will consider a simplified case with massless fermions, for which the corresponding chiral kinetic theory (CKT) in curved space is known \cite{Liu:2018xip}. Focusing on spin response to hydrodynamic gradient, we find that metric perturbation to equilibrium state does formally capture spin response of general hydrodynamic gradient. However with kinematic restrictions by the scattering picture, the vorticity and shear cannot be distinguished. Fortunately this is not the end of the story. In fact the equilibrium state in curved space can be mapped to equilibrium state in flat space with a freedom in the mapping corresponding to a frame choice. A suitable frame choice also allows us to study spin response to temperature gradient, vorticity and shear separately.

We then classify three types of off-equilibrium effect on spin polarization into modifications of (i) spectral function; (ii) distribution function; (iii) Kubo-Martin-Schwinger relation. The last two have been extensively studied using CKT framework. The first one is usually ignore in CKT with free particle being the DOF. The effect of the first type shows up in radiative correction. In this work, we consider radiative correction to spectral function for quarks. It arises from one-loop quark self-energy diagram, whose off-equilibrium correction enters through quark and gluon in the loop. We find the off-equilibrium correction leads spin polarization already at the level of spectral function. This novel contribution to spin polarization lifts the degeneracy in spin responses to vorticity, shear and temperature gradient. This is in close analogy with the counterpart for external electromagnetic fields \cite{Lin:2023egg}. In fact, the story of off-equilibrium effect is more complicated. Since the spectral function is extracted from the retarded function, which resums the self-energy. Apart from off-equilibrium correction to self-energy, the resummation relation itself receives off-equilibrium correction, leading to an extra contribution with local equilibrium self-energy. This effect is studied by solving Kadanoff-Baym equation for retarded function up to next order in gradient. The corresponding contribution turns out to vanish in our case.

The paper is organized as follows: in Sec.~\ref{sec_CKT_curved} we review the CKT in curved space for completeness. We also identify two contributions to spin polarization: one through affine connection unique to curved space; the other is genuine off-equilibrium effect present also in flat space; in Sec.~\ref{sec_vertex}, We show that the effect of metric perturbation in changing particle motion and the definition of correlation function can reproduce the affine connection contribution. It also contains extra contribution that reproduces spin polarization in off-equilibrium state with vorticity only; in Sec.~\ref{sec_offeq}, we illustrate how metric perturbations can be used to mimic off-equilibrium state in the hydrodynamic description. We also discuss the mapping between equilibrium state in curved space and off-equilibrium state in flat space and show how a suitable frame choice allows to study off-equilibrium state with more general hydrodynamic gradient; in Sec.~\ref{sec_radiative} we consider radiative correction to quark spectral function using the off-equilibrium propagators. We will find the radiative correction leads to polarized spectral function. It also lifts the degeneracy in spin responses to vorticity, shear and temperature gradient at tree-level; Sec.~\ref{sec_outlook} is devoted to outlook.

\section{Chiral kinetic theory in curved space}\label{sec_CKT_curved}

In this section, we review the chiral kinetic theory in curved space \cite{Liu:2018xip}. 
The central quantity is the Wigner function (we take lesser two-point function as an example) in a non-trivial metric defined as \cite{Fonarev:1993ht}
\begin{align}\label{S^<_curved}
    S_{\alpha \beta}^<(X=\frac{x_1+x_2}{2},P)=\int d^4y \sqrt{-g(X)} e^{-ip \cdot y}\langle \bar{\psi }_\beta\(X,\frac{y}{2}\) \otimes \psi _\alpha\(X,-\frac{y}{2}\) \rangle=W_{\a\b}(X,P),
\end{align}
where $\psi$ is fermion field. $X=\frac{1}{2}(x_1+x_2)$, $y=x_1-x_2$ being the center and distance of $x_1$ and $x_2$. $\psi (X,-\frac{y}{2})=e^{-\frac{y}{2}\cdot D}\psi(X)$ and $\bar{\psi}(X,\frac{y}{2})=\bar{\psi}(X)e^{\frac{y}{2}\cdot \overleftarrow{D}}$ are fermion fields at $x_1$ and $x_2$ obtained from translating $\ps(X)$ by $-y/2$ and $y/2$ respectively. $\bar{\ps}$ is defined using flat space gamma matrix as $\ps=\ps^\dg\g^{\hat{0}}$ and $\bar{\ps}\overleftarrow{O}=[O\ps]^\dg\g^{\hat{0}}$.\footnote{$\g^{\hat{0}}$ corresponds to the flat space gamma matrix.} The translational operator $D_\mu$ is defined as:
\begin{align}
    D_\mu = \nabla _\mu  + \Gamma _{\mu\nu}^\rho y^\nu \partial _\rho ^y.
\end{align}
The first term on the right-hand side (RHS) is a locally defined covariant derivative, whose action on spinor reads $\nabla_\m\ps=\pd_\m\ps+\frac{1}{4}\o_{\m,ab}\s^{ab}\ps$ with $\sigma ^{ab}=\frac{i}{2}\left[\gamma^a, \gamma^b\right]$.\footnote{In this paper we refer to Greek indices as "curved" and Latin indices as "flat".} The second term that is proportional to the affine connection is called the horizontal lift on the tangent bundle \cite{Fonarev:1993ht}. The horizon lift is introduced such that the phase space variables (chosen to be $y^\n$ and the conjugate momentum $p_\n$) are invariant under the translation
\begin{align}\label{translation_invariance}
D_\m y^\n=\pd_\m^X y^\n-\G_{\m\l}^\n y^\l+\G_{\m\l}^\r y^\l\pd_\r^y y^\n=0.
\end{align}
The commutativity of $D_\m$ and $y^\n$ allows us to exponentialize the translation as
\begin{align}
    \psi (X,-\frac{y}{2})&=\(1-\frac{y^\mu}{2} D_\mu + \frac{y^\mu}{2}\frac{y^\nu}{2} D_\mu D_\nu+...\)\psi(X) \nonumber \\
&=e^{-\frac{y}{2}\cdot D}\ps(X).
\end{align}
The exponential is the gravitational analog of gauge link for fermions in external electromagnetic fields. We shall simply refer to it as gauge link. Counting the phase space variables as $O(\pd^0)$, we find the gravitational gauge link is inherently $O(\pd)$. 

Now we derive the chiral kinetic equation using the Dirac equation $\g^\m\nabla_\m\ps(X)=0$. 
We start with
\begin{align}
    &\partial^y_\mu \left\langle \bar{\psi }_\beta\(X,\frac{y}{2}\) \otimes \psi _\alpha\(X,-\frac{y}{2}\) \right\rangle = \left\langle \partial^y_\mu \bar{\psi }_\beta\(X,\frac{y}{2}\) \otimes \psi _\alpha\(X,-\frac{y}{2}\) + \bar{\psi }_\beta\(X,\frac{y}{2}\) \otimes \partial^y_\mu\psi _\alpha\(X,-\frac{y}{2}\) \right\rangle \nonumber\\
    &= \left\langle \bar{\psi }_\beta\(X,\frac{y}{2}\) \frac{1}{2}\overleftarrow{D}_\mu \otimes \psi _\alpha\(X,-\frac{y}{2}\) + \bar{\psi }_\beta\(X,\frac{y}{2}\) \otimes \(-\frac{1}{2}\) D_\mu\psi _\alpha\(X,-\frac{y}{2}\) \right\rangle \label{partial_rho0}
\end{align}
The following relation has been used in the last equality
\begin{align}
D_\mu \psi(x,y)&=\(\nabla_\mu-\Gamma^\lambda_{\mu\nu}y^\nu \partial^y_\lambda \)\psi(x,y) \nonumber\\
&=e^{y\cdot D}D_\mu \psi+\mathcal{O}(\partial^2) \nonumber\\
&=\partial^y_\mu \psi(x,y)+\mathcal{O}(\partial^2), \label{D_psi}
\end{align}
By contracting with $\gamma^\mu$, (\ref{partial_rho0}) can be written as 
\begin{align}
    &\slashed{\partial}_\m^y \left\langle \bar{\psi }_\beta\(X,\frac{y}{2}\) \otimes \psi _\alpha\(X,-\frac{y}{2}\) \right\rangle \nonumber\\
    &= \frac{1}{2}\g^\m D_\m \left\langle \bar{\psi }_\beta\(X,\frac{y}{2}\) \otimes \psi _\alpha\(X,-\frac{y}{2}\) \right\rangle - \left\langle \bar{\psi }_\beta\(X,\frac{y}{2}\) \otimes \g^\m D_\m \psi _\alpha\(X,-\frac{y}{2}\) \right\rangle. \label{partial_rho}
\end{align}
Using (\ref{D_psi}), we can show that the second term vanishes by the Dirac equation:
\begin{align}
    \g^\m D_\m\psi(X,-\frac{y}{2})&=\g^\m e^{-\frac{y}{2}\cdot D} D_\m\psi(X)+ \mathcal{O}(\partial^2) \nonumber\\
    &=[\g^\m,e^{-\frac{y}{2}\cdot D}]D_\m\psi(X)+e^{-\frac{y}{2}\cdot D}\g^\m D_\m\psi(X)=\mathcal{O}(\pd^2).
\end{align}
Since $\g^\m=e^\m_a(X)\g^a$ does not depend on phase space variable, we may replace $D$ in the commutator in the first term by $\nabla$. It is then easy to see the first term vanishes by vielbein postulate $\nabla_\m e^\n_a=0$. For the second term, we can also replace $D_\m\ps(X)$ by $\nabla_\m\ps(X)$, which vanishes by Dirac equation.
%
Dropping the second term on the RHS of \eqref{partial_rho} and performing the Fourier transform $\int d^4y\sqrt{-g(X)}e^{-ip\cdot y}$ on two sides, we obtain
\begin{align}
    e^\mu_a \gamma^a\(p_\mu+\frac{i}{2}D^\mu\)W(X,P)=0 \label{partial_rho_complete}
\end{align}
where the covariant derivative $D_\mu$ converts into the following form on cotangent bundle:
\begin{align}
    D_\mu=\nabla_\mu +\Gamma^\lambda_{\mu\nu} p_\lambda \partial ^\nu_p.
\end{align}
\eqref{partial_rho_complete} can be decomposed into component form. In the massless theory, the Wigner function contains vector and axial components only, which is decomposed as
\begin{align}
    W(X,P)=\frac{1}{4}\(\gamma^a V_a(X,P) + \gamma^5 \gamma^a A_a(X,P)\).
\end{align}
We use flat space gamma matrices for the decomposition of Wigner function to be consistent with the definition of $\bar{\ps}$, where flat space gamma matrix is used\footnote{In \cite{Liu:2018xip}, curved space gamma matrices are used for the decomposition instead.}. 
Plugging it into (\ref{partial_rho_complete}), we obtain the following explicit expressions:
\begin{align}
    e^\mu_a \gamma^a \[\(p_\mu+\frac{i}{2}\bar{D}_\mu\)\(\gamma^b V_b+\gamma^5 \gamma^b A_b\)+\frac{i}{4}\omega_{\mu cd}\(\(-V^c \gamma^d+V^d \gamma^c\)+\gamma^5\(-A^c\gamma^d+A^d\gamma^c\) \)\]=0
\end{align}
where $\bar{D}_\mu=\partial _\mu +\Gamma^\lambda_{\mu\nu} p_\lambda \partial ^\nu_p$ is the spin-independent part of the covariant derivative $D_\mu$.
By taking the traces with $1$, $\g^5$ and $\g^c\g^d$ respectively, we can derive the equations that the components satisfy:
\begin{subequations}\label{EOM0}
\begin{align}
    &e^\mu_a\left[ \left(p_\mu+\frac{i}{2}\bar{D}_\mu\right)V^a + \frac{i}{4}\omega_{\mu cd} \left(-V^c \eta^{da}+V^d \eta^{ca}\right) \right]=0; \\
    &e^\mu_a\left[ \left(p_\mu+\frac{i}{2}\bar{D}_\mu\right)A^a + \frac{i}{4}\omega_{\mu cd} \left(-A^c \eta^{da}+A^d \eta^{ca}\right) \right]=0; \\
    &e^\mu_a p_\mu V_b\left(-\eta^{ca}\eta^{db}+\eta^{cb}\eta^{da}\right)-\frac{1}{2}e^\mu_a \partial_\mu \epsilon^{cdab}A_b-\frac{1}{4}e^\mu_a \epsilon^{cdab}\left(-\omega_{\mu eb}+\omega_{\mu be}\right)A^e=0. 
\end{align}
\end{subequations}
We can further rewrite the equations above with left/right-handed components defined as $R_a=(V_a+A_a)/2$ and $L_a=(V_a-A_a)/2$. Keeping only the equations for right-handed component, we have
\begin{subequations}\label{eom_Wigner}
	\begin{gather}
	p_a R^a=0;\\
	e^\mu_a D_\mu R^a=0;\\
	-p^c R^d+ p^d R^c - \frac{1}{2}e^\mu_a D_\mu \epsilon^{cdab}R_b=0.
	\end{gather}
\end{subequations}
It is crucial that we decompose the Wigner function using flat gamma matrices, so that the hermiticity of the Wigner function in collisionless theory is manifested as reality of the components, which allows us to separate the real and imaginary parts of the equations in \eqref{EOM0}\footnote{Explicitly, this is due to the hermitian property of the flat space gamma matrices $\g^{\hat{0}}\g^a{}^\dg\g^{\hat{0}}=\g^a$.}. In flat space limit, we have $e^\m_a D_\m\to\pd_a$, \eqref{eom_Wigner} reduces to counterpart in \cite{Hidaka:2016yjf}. In fact we can replace $\pd_a\to e^\m_a D_\m$ in the corresponding solution in \cite{Hidaka:2016yjf} and show that the resulting solution below indeed satisfies \eqref{eom_Wigner}.
\begin{align}
    R^a=4\p \(\d(P^2)p^a f + \frac{\epsilon^{abcd}n_d }{2p\cdot n}e^\mu_b D_\mu \left[p_c \delta(P^2)f\right]\).
\end{align}
Here $n$ is an arbitrary frame vector. Using the commutativity of $D_\m$ with $p_\n$ and $e^\n_a$, we can further simplify the solution above as
\begin{align}
    R^a=4\pi\(p^a\d(P^2) f + \frac{\epsilon^{abcd}n_d }{2p\cdot n}e^\mu_b p_c \delta(P^2) D_\mu f\).\label{R_mu_curved}
\end{align}
For off-equilibrium state characterized by hydrodynamic gradient, we may take the frame vector to be the fluid velocity vector $n^\m=u^\m$ \cite{Hidaka:2017auj} and take a local equilibrium distribution $f[\b(X)p^a u_a(X)]$ with $\b$ being the inverse temperature to rewrite $D_\m f$ as
\begin{align}
D_\mu f&= f'\(\b p^\n\partial_\mu u_\nu +\pd_\m\b p^\n u_\n + \Gamma^\nu_{\mu\lambda}p_\n u^\lambda\).\label{D_mu f}
\end{align}
The first two terms characterize off-equilibrium effect, which also exists in flat space. The last term is a curvature effect. We will scrutinize in the next section on the equivalence between hydrodynamic off-equilibrium field theory and equilibrium field theory with metric perturbation. We will also show in Sec.~\ref{sec_offeq} how a suitable choice of equilibrium state in curved space allows us to access off-equilibrium state in flat space.

\section{Particle motion in metric fields}\label{sec_vertex}

In this section, we treat the curved space as a background metric field, and consider its modification of particle motion. This is an alternative approach to chiral kinetic theory. In case of background electromagnetic field, the equivalence of the two approaches have been shown \cite{Lin:2023egg}. We will see the equivalence is not straightforward in case of metric field. There are two types of contributions: on one hand, the metric explicitly change the particle motion by scattering with the particles; on the other hand, the metric implicitly modifies the definition of Wigner function via the gauge link. Both contributions are at least $O(\pd)$. The former is because metric couples to spinor through spin connection, which is $O(\pd)$. The latter has been seen explicitly in the previous section. Since $\pd_X$ is nonvanishing only on curved metric, we can expand the metric as
%
\begin{align}
    g_{\mu\nu}\simeq \eta_{\mu\nu}+ h_{\mu\nu},
\end{align}
and work up to first order in the perturbation $h_{\mu\nu}$. The vielbein can be expanded accordingly as $e^\mu_a \simeq e^\mu_{a(0)} + \delta e^\mu_a$, with 
\begin{gather}\label{vielbein}
e_{a(0)}^\m e_{b(0)}^\n\h_{\m\n}=\h_{ab},\nonumber\\
e_{a(0)}^\m \d e_{b}^\n \h_{\m\n}+\d e_{a}^\m e_{b(0)}^\n\h_{\m\n}=-e_{a(0)}^\m e_{b(0)}^\n h_{\m\n}.
\end{gather}
%
%
%
%
By assuming the spacetime is torsionless and the condition that $\nabla _\rho g_{\mu\nu}=0$, the corresponding affine connection and spin connection can be given by
\begin{align}
    \Gamma^{(0)\lambda}_{\mu\nu}&=\omega^{(0)}_{\mu, ab}=0;\\
    \Gamma^{(1)\lambda}_{\mu\nu}&=\frac{1}{2}\eta^{\lambda\sigma}\left(h_{\sigma\mu,\nu}+h_ {\nu\sigma,\mu}-h_{\mu\nu,\sigma}\right);\\
    \omega^{(1)}_{\mu, ab}&=\eta_{\alpha \beta}{e^\alpha_a}_{(0)}\left( \partial_\mu \delta e^\beta_b + \frac{1}{2}\eta^{\beta\sigma}\left(h_{\sigma\mu,\nu}+h_ {\nu\sigma,\mu}-h_{\mu\nu,\sigma}\right){e^\nu_b}_{(0)}\right) \nonumber\\
    &=\eta_{\alpha \beta}{e^\alpha_a}_{(0)} \partial_\mu \delta e^\beta_b+\frac{1}{2}{e^\sigma_a}_{(0)} {e^\nu_b}_{(0)}\[ h_{\mu \sigma ,\nu}-h_{\mu \nu ,\sigma} \],\label{spin_connection_linear}
\end{align}
where the subscripts $(0)$ and $(1)$ represent the orders in $\pd_X$.

We pause to discuss property of different quantities under local Lorentz transformation. For a given metric field $g_{\m\n}$, it is possible to choose different vielbein field $e_a^\m$ satisfying \eqref{vielbein}, with different choices related by Lorentz transformation. The infinitesimal Lorentz transformation parameterized by $\th_{ab}(X)$ acts on different quantities as follows
\begin{align}\label{lLT}
&\ps\to \(1-\frac{i}{4}\th_{ab}\s^{ab}\)\ps,\nonumber\\
&e^a_\m\to e^a_\m+\th^a{}_b e^b_\m,\nonumber\\
&\o_\m^{ab}\to\o_\m^{ab}-\(\pd_\m\th^{ab}+\o_\m^{a}{}_c\th^{cb}+\o_\m^{b}{}_{c}\th^{ac}\).
\end{align}
The spin connection is introduced to the covariant derivative such that it transforms as
\begin{align}
\nabla_\m\ps\to (1-\frac{i}{4}\th_{ab}\s^{ab})\nabla_\m\ps.
\end{align}
In particular terms containing $\pd_\m\th_{ab}$ cancel in the transformations of $\pd_\m\ps$ and $\frac{1}{4}\o_{\m,ab}\s^{ab}\ps$ in the above.
Specializing to our case, we choose constant $e_{a(0)}^\m$ and spacetime dependent $\d e_a^\m$. In this case, only the first term on the RHS of \eqref{spin_connection_linear} gives rise to a contribution dependent on $\pd_\m\th_{ab}(X)$, which is to be canceled by the counterpart from the transformation of $\pd_\m\ps$, thus does not appear in the Wigner function. This allows us to drop the first term and at the same time do not consider local Lorentz transformation on $\ps$.
%

Now we are ready to calculate the gauge link contribution to the Wigner function keeping only the second term on the RHS of \eqref{spin_connection_linear} for the spin connection. Expanding the spinor in the Wigner function up to $O(\pd)$, we find
\begin{align}
\ps(X,-\frac{y}{2})=\ps(X)-\frac{y^\m}{2}\(\pd_\m^X+\frac{1}{4}\o_{\m,ab}\s^{ab}+\G_{\m\n}^\r y^\n\pd_\r^y\)\ps(X),
\end{align}
and similarly for $\bar{\ps}(X,\frac{y}{2})$.
At this order the affine connection term does not contribute. The partial derivative term could contribute if we consider local Lorentz transformation on $\ps$. As we argued above, we can ignore it together with the first term on the RHS of \eqref{spin_connection_linear}. Keeping only the spin connection term, we have
\begin{align}\label{first_order_S<}
    S_{\a\b}^{<(1)}(x_1,x_2)&=\frac{y^\mu}{2}\Bigg\langle \(\left(-\frac{i}{4}\omega_{\mu,ab}\sigma^{ab} \psi(X)\right)^\dagger\gamma^0\)_\b \psi_\alpha(X) 
    - \bar{\psi}_\beta(X) \left(-\frac{i}{4}\omega_{\mu,ab}\sigma^{ab}\right)\psi_\alpha(X)\Bigg\rangle,
\end{align}
Proceeding with the calculations by restricting to particles with positive energy and introducing the plane-wave expansion, we have
\begin{align}
    S^{<(1)}(p_1,p_2)&= \frac{y^\mu}{2}\int_{p_1,p_2} e^{-ip_2 \cdot x_2+ip_1 \cdot x_1}\frac{1}{2\sqrt{p_1 p_2}}\frac{i}{4}\omega_{\mu,ab}\sum_{r,s}\Big[u^s_\alpha(p_1)\left((\sigma^{ab}u^r(p_2))^\dagger\gamma^0\right)_\beta \nonumber\\
    &+\left(\sigma^{ab}u^s(p_1)\right)_\alpha\bar{u}^r_\beta(p_2) \Big]\langle a^{r\dagger}_{p_2}a^s_{p_1}\rangle \nonumber\\
    &=\frac{y^\mu}{2}\int_{p'} e^{ip' \cdot y}\frac{1}{2p'}\frac{i}{4}\omega_{\mu,ab}\Big[\(\s^{ab}\slashed{P'}\)_{\a\b}+\(\slashed{P'}\s^{ab}\)_{\a\b}\Big]f(p')
\end{align}
where $\int_{p_i}=\int \frac{d^3 p_i}{(2\pi)^3}$ and $p_i=|\vec{p}_i|$. 
In the last line we have taken the expectation value in an equilibrium state $\langle a^{r\dagger}_{p_2}a^s_{p_1}\rangle=(2\pi)^3 \delta^{(3)}(\vec{p}_2-\vec{p}_1)f(p_1)\d^{rs}$ following the spirit of mimicking off-equilibrium effect by metric perturbation of equilibrium state encoded in $\o_{\m,ab}$ above. We have also used $\sum_s\left(\sigma^{ab}u^s(p')\right)_\alpha\bar{u}^s_\beta(p')=\left(\sigma^{ab}\slashed{P'} \right)_{\alpha\beta}$ and $\sum_s u^s_\alpha(p')\left((\sigma^{ab}u^s(p'))^\dagger\gamma^0\right)_\beta=\left(\slashed{P'}\sigma^{ab}\right)_{\alpha \beta}$ for on-shell momentum $p'$.

The gauge link contribution is to be combined with the other one corresponding to fermion scattering on background metric field. The combined result will be shown to agree with the counterpart of CKT \eqref{R_mu_curved}. The fermion scattering on background metric field is most conveniently described by field theory in $ra$-basis. We shall show the agreement for $S_{rr}$ instead.
The latter is related to $S^{<}$ by the following identity 
\begin{align}
    S^{rr}(x_1,x_2)=\frac{1}{2}\left( S^>(x_1,x_2)-S^<(x_1,x_2) \right).
\end{align}
$S^{>(1)}$ can be obtained from $S^{<(1)}$ by simply commuting the creation and annihilation operators, which amounts to using $\langle a^s_{p_1}a^{r\dagger}_{p_2}\rangle=(2\pi)^3 \delta^{(3)}(\vec{p}_2-\vec{p}_1)(1-f(p_1))\d^{rs}$. We then have for the Wigner transformed $S^{rr}$
\begin{align}\label{Srr_exp}
    S_{\a\b}^{rr}(X,P)=&\int d^4y e^{-ip \cdot y} \frac{y^\mu}{2}\int_{p'} e^{ip' \cdot y}\frac{1}{2 p'}\frac{i}{4}\omega_{\mu,ab}\Big[\(\s^{ab}\slashed{P'}\)_{\a\b}+\(\slashed{P'}\s^{ab}\)_{\a\b}\Big]\(\frac{1}{2}-f(p')\)\nonumber\\
    =&2\pi \frac{i\partial^\mu_p}{2}\[\delta(p_0 ^2-p ^2)\theta(p_0)\frac{i}{4}\omega_{\mu,ab}\(\sigma^{ab}\slashed{P}+\slashed{P}\sigma^{ab}\)_{\alpha \beta}\(\frac{1}{2}-f(p)\)\].
\end{align}
We can drop $\th(p_0)$ as we consider particle with positive energy. It is instructive to inspect the Lorentz transformation of the expression above. With \eqref{spin_connection_linear}, $S_{\a\b}^{rr}$ transforms as a tensor following the transformation of $\o_{\m,ab}$ in \eqref{lLT}. This is consistent with the transformation of $S^{rr}$ as a bispinor $S^{rr}\to \L_{1/2}S^{rr}\L_{1/2}^{-1}$ with $\L_{1/2}=\exp\(-\frac{i}{4}\th_{ab}\s^{ab}\)$. \eqref{Srr_exp} is to be combined with the other contribution from particle scattering with metric field, which we shall see shortly is independent of choice of vielbein $e_{a(0)}^\m$. Therefore we need to fix the ambiguity by choosing a Lorentz frame. A natural choice is $e_{a(0)}^\m=\d^\m_a$.
This amounts to choosing the medium frame in the flat space limit. It leads to the following right-handed component
\begin{align}\label{R_rr^lambda}
    \mathcal{R}_{rr}^c=&\frac{1}{2}\tr\big[\g^c(1+\g^5)S_{rr}\big]\nonumber\\
    =&\pi\epsilon^{cabd} \(-2p_d p^\mu\delta'(P^2)\) \(f(p)-\frac{1}{2}\) \frac{1}{2}(h_{\mu a,b}-h_{\mu b,a}) \nonumber\\
    &+\pi \epsilon^{cabd}p_d \delta(P^2) f'(p)\delta^{\mu i}\frac{p_i}{p}\frac{1}{2}(h_{\mu a,b}-h_{\mu b,a}).
\end{align}

\begin{figure}
	\centering
	\includegraphics[width=.4\textwidth]{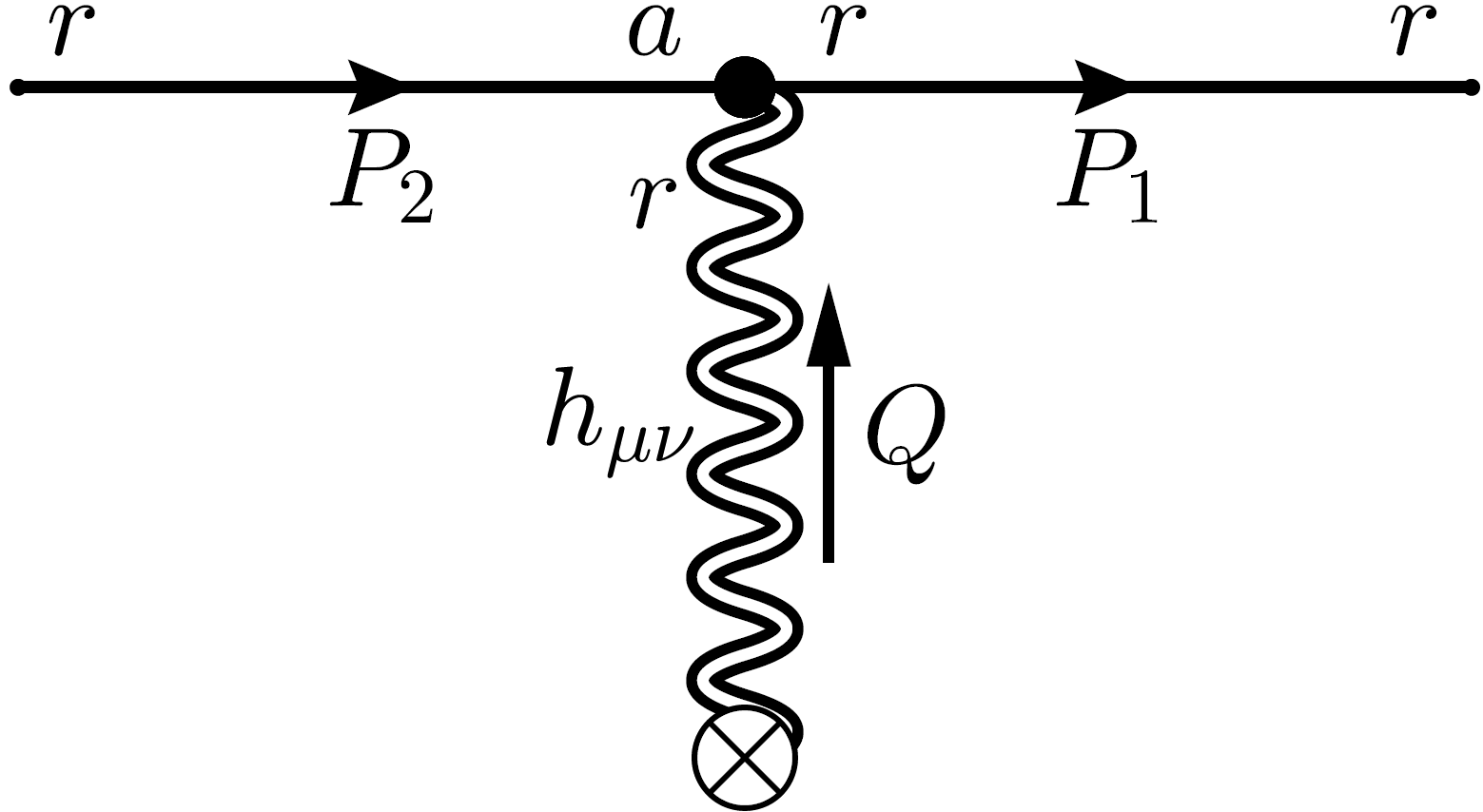}
        \includegraphics[width=.4\textwidth]{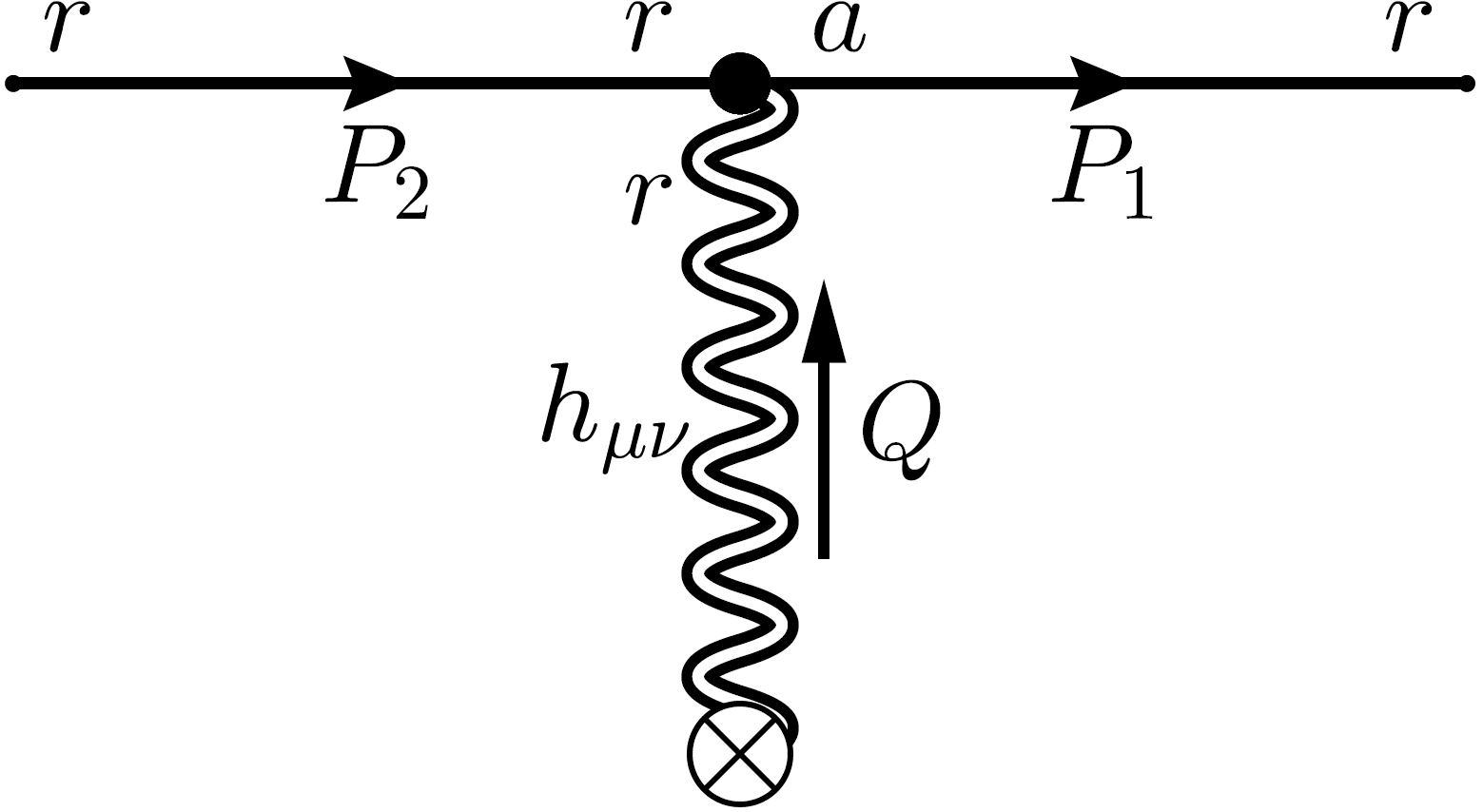}
	\caption{massless fermions scattering in background metric field. Assume that the momentum of the metric perturbation $q$ is very small: $|\vec{q}|\ll|\vec{p}|$}.
        \label{fig:tree_level_diagrams}
\end{figure}

Now we work out the contribution from particle scattering with metric field. Fig.~\ref{fig:tree_level_diagrams} show the relevant diagrams in $ra$-basis with the background metric field labeled by $r$-field. The propagators in $ra$-basis are given by
\begin{align}
&S_{ra}(Q)=\frac{i{\slashed Q}}{(q_0+i\h)^2-\vec{q}^2},\nonumber\\
&S_{ar}(Q)=\frac{i{\slashed Q}}{(q_0-i\h)^2-\vec{q}^2},\nonumber\\
&S_{rr}(Q)=-2\p\e(q_0)\d(Q^2){\slashed Q}\(f(q_0)-\frac{1}{2}\),
\end{align}
Explicit expression of the sum of the two diagrams are given by
\begin{align}\label{Srr_diag}
S^{rr}=&-2\pi \Bigg[\epsilon(p_1^0)\delta(P_1^2) \(f(p_1^0)-\frac{1}{2}\)\slashed{P}_1\gamma^{\{a}P^{b\}}\frac{i\slashed{P}_2}{P_2^2+i\e(p_2^0)\h} \nonumber\\
&+\frac{i\slashed{P}_1}{P_1^2-i\e(p_1^0)\h}\gamma^{\{a}P^{b\}}\epsilon(p_2^0)\delta(P_2^2) \slashed{P}_2\(f(p_2^0)-\frac{1}{2}\)\Bigg]\frac{i}{2}h_{ab}, 
\end{align}
with $P=\frac{1}{2}(P_1+P_2)$ and $Q=P_1-P_2$ corresponds to the momentum exchange between particle and metric field. Clearly this expression depends on metric only. This is because the rest frame of the medium is implicitly chosen as is clear from the expression of the propagators.

We need some kinematic assumption to proceed: \eqref{Srr_diag} indicates that either $P_1$ or $P_2$ is on-shell. We are interested in the regime where both are on-shell so that the scattering picture is meaningful. This leads to the following constraint
\begin{align}\label{nwc1}
\(P\pm\frac{Q}{2}\)^2=0\to P\cdot Q=0.
\end{align}
On the other hand, $q_0$ corresponds to energy exchange between particle and metric field. A non-vanishing $q_0$ would disturb the equilibrium distribution. So we further require
\begin{align}\label{nwc2}
q_0=0.
\end{align}
\eqref{nwc1} and \eqref{nwc2} are the gravitational analog of no-work condition for fermions in background electromagnetic fields \cite{Lin:2023egg}. These conditions allow us to replace $P_i^2\to P^2$ (up to $\mathcal{O}(Q^2)\sim\mathcal{O}(\pd_X^2)$ ) and $p_i^0\to p_0$. We can further drop the sign function and symmetrization to have
\begin{align}
S^{rr}=-2\pi \[\delta(P^2) \(f(p_0)-\frac{1}{2}\)\slashed{P}_1\gamma^{a}P^{b}i\slashed{P}_2\(\frac{1}{P^2+i\h}+\frac{1}{P^2-i\h}\)\]h_{ab}.
\end{align}
We then extract the right-handed component as in \eqref{R_rr^lambda}. Using
\begin{align}\label{trace}
\tr\[\g^c(1+\g^5)\slashed{P}_1\g^a\slashed{P}_2\]\stackrel{\mathcal{O}(Q)}{=}4i\e^{cdae}p_d q_e,
\end{align}
in which only contribution at $\mathcal{O}(Q)$ is kept, we have
\begin{align}\label{Rrr_diag}
\mathcal{R}_{rr}^c=2\p\d'(P^2)\(f(p_0)-\frac{1}{2}\)(-2\e^{cdae})p_dq_ep^b\frac{i}{2}h_{ab}.
\end{align}
$\d'(P^2)$ arises from using the following relation
\begin{align}
\frac{1}{x\pm i\e}\d(x)=-\frac{1}{2}\d'(x)\mp\p\(\d(x)\)^2.
\end{align}

Now we are ready to combine \eqref{R_rr^lambda} and \eqref{Rrr_diag}. Using that $iq_e h_{ab}$ is nothing but the Fourier transform of $h_{ab,e}$, we find the first term of \eqref{R_rr^lambda} and \eqref{Rrr_diag} look almost the same: apart from the overall sign, the only difference is the argument of $f$. With $\d'(P^2)$ in the prefactor, the . Using
\begin{align}
\d'(P^2)\(f(p_0)-f(p)\)=-\frac{1}{2p}\d(P^2)f'(p),
\end{align}
we arrive at the following contribution after partial cancellation of the first term of \eqref{R_rr^lambda} and \eqref{Rrr_diag}
\begin{align}
2\p\d(P^2)\e^{cdae}\frac{p_dp^b}{2p}\d(P^2)f'(p)h_{ba,e},
\end{align}
to be combined with the second term of \eqref{R_rr^lambda}. To write out explicit components of the combined contribution, we specify the components of metric perturbation: Since we consider off-equilibrium state characterized by hydrodynamic gradient, which couples to $h_{00}$ and $h_{0i}$ only, we turn on those components of $h_{ab}$. This leads to the following explicit components
\begin{align}\label{Rrr_combined}
&R^0=4\p\d(P^2)f'(p)\big[\frac{1}{4}\e^{ijk}p_kh_{0i,j}\big],\nonumber\\
&R^i=4\p\d(P^2)f'(p)\big[-\frac{1}{4}\e^{ijk}p_0 h_{0j,k}+\frac{1}{2}\e^{ijk}\frac{p_kp_l}{p}h_{l0,j}+\frac{1}{4}\e^{ijk}p_kh_{00,j}\big]
\end{align}
We shall compare \eqref{Rrr_combined} with the affine connection term of \eqref{R_mu_curved}. They correspond to effect of metric perturbation on equilibrium state from different approaches, which we hope can mimic off-equilibrium state effect.
We spell out the affine connection term of \eqref{R_mu_curved} (see \eqref{D_mu f}),
\begin{align}\label{Rrr_CKT}
R^i&=4\p\frac{\e^{ijk}p_k}{2p}\d(P^2)f'(p)\(\G_{j0}^0p_0+\G_{j0}^l(-p_l)\)\nonumber\\
&=4\p\frac{\e^{ijk}p_k}{2p}\d(P^2)f'(p)\(p_0\frac{1}{2}h_{00,j}+p_l\frac{1}{2}\(h_{0l,j}-h_{0j,l}\)\).
\end{align}
We first note that the last term in \eqref{Rrr_CKT} vanishes by our kinematic conditions \eqref{nwc1} and \eqref{nwc2} as: $p_lh_{0j,l}\sim p_lq_l=p_0q_0-P\cdot Q=0$. We then find the $h_{00,j}$ term in \eqref{Rrr_CKT} reproduces the counterpart in \eqref{Rrr_combined}, whileas the $h_{0l,j}$ term only reproduces half of the counterpart in \eqref{Rrr_combined}. The extra terms in \eqref{Rrr_combined} are given by
\begin{align}\label{spin_vorticity}
&R^0=4\p\d(P^2)f'(p)\big[\frac{1}{4}\e^{ijk}p_kh_{0i,j}\big],\nonumber\\
&R^i=4\p\d(P^2)f'(p)\big[-\frac{1}{4}\e^{ijk}p_0 h_{0j,k}+\frac{1}{4}\e^{ijk}\frac{p_kp_l}{p}h_{l0,j}\big].
\end{align}
Interestingly, we can identify them as the contribution in the presence of vorticity \cite{Hidaka:2017auj,Gao:2018jsi}. To see that, we set
\begin{align}\label{huT}
	h_{0i}=-u_i,\quad \frac{1}{2}h_{00}=-\frac{\d T}{T}.
\end{align} 
(this is known in hydrostatic theory in curved space \cite{Banerjee:2012iz} and will be further justified in the next section) and use the following relations
\begin{align}\label{vorticity_relation}
&\e^{ijk}h_{0i,j}=\e^{ijk}\pd_ju_i=-2\o^k,\nonumber\\
&p_l\e^{ijk}h_{l0,j}=p_l\e^{ijk}\(h_{l0,j}-h_{j0,l}\)=2p_l\e^{ijk}\e^{ljm}\o^m,
\end{align}
with $\o^i=-\frac{1}{2}\e^{ijk}\pd_j u_k$ being the vorticity. The introduction of the vanishing $h_{j0,l}$ term uses again the kinematic condition.

In summary, we have found the picture of particle scattering in metric fields gives rise to to two contributions to the Wigner function: the first one is the affine connection contribution unique to curved space; the second is the response to vorticity, which is a particular off-equilibrium effect in flat space we need. The results have been obtained under two kinematic constraints: (i) no energy exchange between particle and medium; (ii) momentum exchange is orthogonal to particle momentum. Let us see if we can use either of these to mimic the general off-equilibrium effect. We flip the sign of the last vanishing term in \eqref{Rrr_CKT} using the kinematic conditions \eqref{nwc1} and \eqref{nwc2} to have
\begin{align}\label{spin_others}
R^i&=4\p\frac{\e^{ijk}p_k}{2p}\d(P^2)f'(p)\(p_0\frac{1}{2}h_{00,j}+p_l\frac{1}{2}\(h_{0l,j}+h_{0j,l}\)\)\nonumber\\
&=4\p\frac{\e^{ijk}p_k}{2p}\d(P^2)f'(p)\(\frac{\pd_j\b}{\b}+ p_l\s_{jl}\),
\end{align}
where we have used \eqref{huT} and $\s_{jl}=-\frac{1}{2}\(\pd_j u_l+\pd_l u_j\)+\frac{1}{3}\d_{jl}\pd_k u_k$ being the shear tensor. We readily identify the above as spin responses to temperature gradient and shear. Together with the spin response to vorticity in \eqref{spin_vorticity}, we find the picture of particle scattering in metric fields does reproduce the spin response to general hydrodynamic gradients\footnote{It is also possible to achieve the same with the CKT solution in curved space \eqref{R_mu_curved} provided that we extend the equilibrium state to the one with vorticity present. In this way we can also have \eqref{spin_vorticity} from the non-affine connection terms with modified distribution in \eqref{R_mu_curved} \cite{Hidaka:2017auj,Gao:2018jsi}.}

The equivalence found above is limited by the kinematic conditions, by which we actually cannot even distinguish vorticity and shear. We could also take $-h_{j0,l}$ instead of $h_{j0,l}$ in \eqref{spin_others} to obtain the vorticity. While the idea of mimicking off-equilibrium effect by metric perturbation on equilibrium state seems to have only limited applicability, we can still work directly with off-equilibrium state. 
In the next section, we will discuss how an off-equilibrium state emerges naturally in the presence of metric perturbation, which allows us to drop \eqref{nwc1}.

\section{Off-equilibrium effect from metric perturbation}\label{sec_offeq}

The off-equilibrium state we consider is characterized by hydrodynamic gradient. The spin polarization can be induced by vorticity, shear, acceleration and temperature gradient in neutral fluid. Since spin is enslaved by momentum in massless case, the spin response to hydrodynamic gradient can be viewed as completely equilibrated. In massive case, spin mode equilibrates independent of momentum \cite{Gao:2019znl,Weickgenannt:2019dks,Hattori:2019ahi,Liu:2020flb,Sheng:2020oqs}, and may couple with hydrodynamic modes \cite{Hattori:2019lfp,Hongo:2021ona}. While turning on slow-varying metric perturbations is expected to excite hydrodynamic modes in general, the actual response can be complicated. The situation simplifies drastically in the static limit, with $h_{00}$ and $h_{0i}$ mimicking temperature variation and fluid velocity respectively. We shall show this more explicitly from hydrodynamic correlation functions. The limitation of the static limit is that the corresponding metric perturbations cannot mimic acceleration, which involves time derivative.

For simplicity, we consider neutral fluid, for which the retarded correlation functions of energy-momentum tensor (EMT) are given by \cite{Kovtun:2012rj}
\begin{align}\label{hydro_retarded}
&G_{\p_i\p_j}^R=\(\d_{ij}-\hat{k}_i\hat{k}_j\)\frac{\h k^2}{i\o-\g_\h k^2}+\hat{k}_i\hat{k}_j\frac{w_0(c_s^2k^2-i\g_s\o k^2)}{\o^2-c_s^2k^2+i\g_s\o k^2},\nonumber\\
&G_{\e\e}^R=\frac{w_0k^2}{\o^2-c_s^2k^2+i\g_s\o k^2}\nonumber\\
&G_{\e\p_i}^R=G_{\p_i\e}^R=\frac{w_0\o k_i}{\o^2-c_s^2k^2+i\g_s\o k^2},
\end{align}
with $\p_i$ and $\e$ denoting momentum density and energy density respectively. $w_0=\e_0+p_0$ is the equilibrium enthalpy. $\h$, $\z$ and $\c_s$ are shear viscosity, bulk viscosity and speed of sound respectively. The damping coefficients are defined as $\g_\h=\frac{\h}{w_0}$, $\g_s=\frac{4\h/3+\z}{w_0}$. In the static limit, we set $\o=0$, \eqref{hydro_retarded} reduces to
\begin{align}\label{hydro_static}
G_{\p_i\p_j}^R=-\d_{ij}w_0,\quad G_{\e\e}^R=-\frac{w_0}{c_s^2}.
\end{align}
Now we consider the following components of metric perturbation $h_{00}$ and $h_{0i}$. From linear response theory, we have
\begin{align}
&\d\p_i=G^R_{\p_i\p_j}h_{0j}=-w_0h_{0i},\nonumber\\
&\d\e=\frac{1}{2}G_{\e\e}^Rh_{00}=-\frac{w_0}{2c_s^2}h_{00}.
\end{align}
In the static limit, momentum density can only be generated by fluid velocity, thus we identify $h_{0i}=-v^i$. On the other hand, energy density can only be generated by temperature variation. Using the thermodynamic relations $w_0=T s$, $dp=sdT$ for neutral fluid and $c_s^2=\frac{\pd p}{\pd \e}$, we find $h_{00}=-2\frac{\d T}{T}$. The interpretation is clear: in the static limit, the hydrodynamic modes equilibrate completely, the response of fluid velocity and temperature to metric perturbations is fixed by thermodynamics. These responses give rise to an off-equilibrium state characterized by static hydrodynamic gradient, which further induces spin polarization. This is the off-equilibrium effect of metric perturbations we referred to in the introduction.

Within CKT in curved space, the off-equilibrium effect can be realized through proper choice of state. It turns out the correct off-equilibrium state nicely corresponds to equilibrium state in curved space. To see this, we need to identify flat space quantities in curved space formulation. A natural choice is to identify momentum with flat indices as flat space momentum, which is related to its curved space counterpart by vielbein $p_a=p_\m e^\m_a$. In the presence of static metric perturbations in $g_{0\m}$ components, the local equilibrium distribution is given by $f(\frac{p_\m u^\m}{T})$ with $u^\m=(g_{00}^{-1/2},0,0,0)$. This can be rewritten in flat space as
\begin{align}
f\(\frac{p_a u^a}{T}\),
\end{align}
where $u^a=u^\m e_\m^a$ and $T_0$ being the flat space temperature. Note that temperature is not invariant under diffeomorphism but scales as $T=T_0 g_{00}^{-1/2}$. Using $g_{00}=1+h_{00}$ and working to linear order in $h_{00}$, we find
\begin{align}\label{dT_h}
\frac{\d T}{T_0}=\frac{T-T_0}{T_0}=-\frac{1}{2}h_{00}.
\end{align}
To find the relation between $u^a$ and $h_{0i}$, we note that the choice of vielbein is not unique. We make the following choice up to linear order in 
\begin{align}
e_0^{\hat{0}}=1+\frac{h_{00}}{2},\quad e_i^{\hat{j}}=-\d_{ij},\quad e_0^{\hat{i}}=-h_{0i}.\label{vielbein_h}
\end{align}
The flat space fluid velocity following from it is given by
\begin{align}
u^{\hat{0}}=1,\quad u^{\hat{i}}=-h_{0i}.
\end{align}
We have confirmed that the equilibrium state in curved space indeed gives rise to to the expected temperature variation and fluid velocity in flat space. Finally we remark that we only impose the static condition on the metric perturbation, which corresponds to the no energy exchange condition \eqref{nwc2} in Sec.~\ref{sec_vertex}. Without \eqref{nwc1}, vorticity and shear can clearly be distinguished.



\section{Radiative correction to quark self-energy and spin polarization}\label{sec_radiative}

With the off-equilibrium effect clarified, we now give general account of the radiative corrections in CKT. We start with the following representation of Wigner function predicted by CKT\footnote{Collisional contribution to the Wigner function is not included here as we are concerned with radiative correction in this work.} \cite{Hidaka:2017auj}.
\begin{align}\label{Wigner_rep}
&S^{<(0)}(P)=\g_\m \r^\m\tilde{f}\nonumber\\
&S^{<(1)}(P)=\g^5\g_\m\big[\d\tilde{f}\r^\m+\frac{\e^{\m\n\r\s}\r_\r u_\s\pd_\n\tilde{f}}{2P\cdot u}\big],
\end{align}
with $\g^\m\r_\m=2\p\,\e(P\cdot u){\slashed P}\d(P^2)$ being the free theory spectral function. \eqref{Wigner_rep} contains two terms in the square bracket: the first term encodes modification of distribution function $\d\tilde{f}$ in an off-equilibrium state. In case of vorticity, $\d\tilde{f}$ is given by spin-vorticity coupling shift of the distribution. The second term arises from free spectral function and local equilibrium distribution, which can be viewed as modification of the KMS relation at $O(\pd)$. The radiative corrections may occur in three types:\\
(i) spectral function. The spectral function is modified by self-energy in equilibrium and its off-equilibrium correction. It can contribute to the axial component of the Wigner function with the standard KMS relation if the spectral function itself develops an axial component. This contribution is proportional to equilibrium distribution.\\
(ii) modified distribution. Ignoring collisional effect, the distribution function is modified due to spin-vorticity coupling only. We have shown in Sec.~\ref{sec_vertex} that it can be described by scattering of particle in background metric field. The corresponding radiative correction can be described by in-medium gravitational form factors \cite{Lin:2023ass}.\\
(iii) modified KMS relation. Even with local equilibrium self-energy correction containing only vector component, the axial component of the Wigner function can still receives contribution in a way similar to second term on the RHS of \eqref{Wigner_rep}. This contribution is always proportional to derivatives of distribution function in contrast to type (i) contribution.\\
Given that the last two contributions exist already at tree-level and are well-understood in the CKT framework. The first contribution is present with radiative correction only. We shall focus on the first one and study its impact on spin polarization, leaving radiative corrections to the other two types for future explorations.

\subsection{Modified spectral function}
The operator definition of spectral function is given by
\begin{align}\label{spec_def}
\r(P)=\int d^4x e^{iP\cdot x}\langle\ps_\a(x)\bar{\ps}_\b(0)+\bar{\ps}_\b(0)\ps_\a(x)\rangle.
\end{align}
A closely related quantity is the retarded function defined as
\begin{align}\label{ra_def}
S_{ra,\a\b}=\int d^4x e^{iP\cdot x}\th(x_0)\langle\ps_\a(x)\bar{\ps}_\b(0)+\bar{\ps}_\b(0)\ps_\a(x)\rangle.
\end{align}
Using the representation of Heaviside function
\begin{align}\label{theta}
\th(x_0)=i\int \frac{dk_0}{2\p}\frac{e^{-i k_0x_0}}{k_0+i\h},
\end{align}
we easily obtain
\begin{align}
S_{ra}=\int\frac{dk_0'}{2\p}\frac{i\r(k_0')}{k_0-k_0'+i\h}.
\end{align}
If $\r$ were real, we could simply take the real part of both sides to obtain
\begin{align}\label{spec_ra}
\r(P)=2\text{Re}[S_{ra}(P)].
\end{align}
With $\r$ being a matrix in Dirac space, the reality condition is replaced by: $\g^0\r(P)^\dg \g^0=\r(P)$, which indeed holds at operator level from the definition \eqref{spec_def}\footnote{To confirm the reality condition at expectation value level, we still need hermitian condition in the density of state, or equivalently the state is invariant under time-reversal. We assume this holds for the off-equilibrium state considered in this paper.}. It follows that \eqref{spec_ra} holds with the real part defined as $\text{Re}A=\frac{1}{2}(A+\g^0A^\dg\g^0)$. Importantly, the relation \eqref{spec_ra} holds in off-equilibrium state as well. It suggests us to perform the self-energy calculations in the $ra$-basis \cite{Dong:2024cbt}.

As remarked earlier, equilibrium self-energy and its off-equilibrium correction will lead to radiative correction to spectral function. Below we derive their contribution to the retarded function based on the following Kadanoff-Baym equation \cite{Lin:2021mvw}
\begin{align}
i{\slashed \pd}_x S_R(x,y)-\int d^4z \S_R(x,z)S_R(z,y)=-\d(x-y).
\end{align}
Applying Wigner transform on both sides and performing gradient expansion, we obtain
\begin{align}\label{KB}
\frac{i}{2}{\slashed \pd}S_R(X,P)+{\slashed P}S_R(X,P)-\(\S_R(X,P)S_R(X,P)+\frac{i}{2}\{\S_R(X,P),S_R(X,P)\}_\text{PB}\)=-1,
\end{align}
where $\{\}_\text{PB}$ is the Poisson bracket defined as
\begin{align}
\{A,B\}_\text{PB}=\pd_P A\cdot\pd_X B-\pd_X A\cdot\pd_P B.
\end{align}
$S_R=i S_{ra}$ and $\S_R$ is the retarded self-energy to be expressed in the $ra$-basis later.

\eqref{KB} is solved in appendix~\ref{sec_app_A} in a double expansion of the gradient and size of the self-energy (counted as $g^2$ from one-loop calculations). Up to $O(g^2\pd)$, the solution is given by
\begin{align}\label{SR_sol}
&S_{R}=S_{R}^\bzero+S_{R}^\bone+\cdots,\\
&S_R^\bzero=-\frac{1}{\slashed P}-\frac{1}{\slashed P}\S_R \frac{1}{\slashed P},\nonumber\\
&S_R^\bone=\g^5\g^\b P^\n T^{\m\l}\e_{\b\l\m\n}\frac{-1}{(P^2)^2},\nonumber
\end{align}
where the superscript denoting the order in $\pd_X$ and $T_{\m\l}=\pd_{[\m}\S^R_{\l]}$. The solution has been obtained by assuming $\S_R$ to be local equilibrium self-energy. It is parity invariant thus contains vector component only $\S_R=\g_\l\S_R^\l$. It depends on coordinate through local temperature and fluid velocity giving rise to nonvanishing $T_{\m\l}$. We will also consider off-equilibrium correction to self-energy below, which we denote as $\d\S_R$. It is not difficult to see $\d\S_R$ leads to the following change to $S_R^\bone$
\begin{align}
S_R^\bone=-\frac{1}{\slashed P}\d\S_R \frac{1}{\slashed P}+\g^5\g^\b P^\n T^{\m\l}\e_{\b\l\m\n}\frac{-1}{(P^2)^2}.
\end{align}
The formal solution cannot be used on the free particle $P^2=0$. In vacuum, $i\e$ prescription is needed to regularize the expression. In the medium, the self-energy $\S_R$ shifts the dispersion relation by a small amount. We shall replace ${\slashed P}$ by ${\slashed P}-\S_R\equiv {\slashed P}_\S$ as
\begin{align}\label{SR_sol2}
S_R^\bone=-\frac{1}{\slashed P_\S}\d\S_R \frac{1}{\slashed P_\S}+\g^5\g^\b P^\n T^{\m\l}\e_{\b\l\m\n}\frac{-1}{(P_\S^2)^2}.
\end{align}
The replacement is done in the denominators but not the numerators, which does not affect our counting scheme.
In the regime of $P\gg T$ we consider below, the quasi-particle gains a small damping rate. Using $S_R=iS_{ra}$ and \eqref{spec_ra}, we obtain
\begin{align}
\r=2\text{Im}\big[S_R\big].
\end{align}

\subsection{Self-energy in equilibrium and off-equilibrium correction}
\begin{figure}
    \centering
    \includegraphics[width=0.4\linewidth]{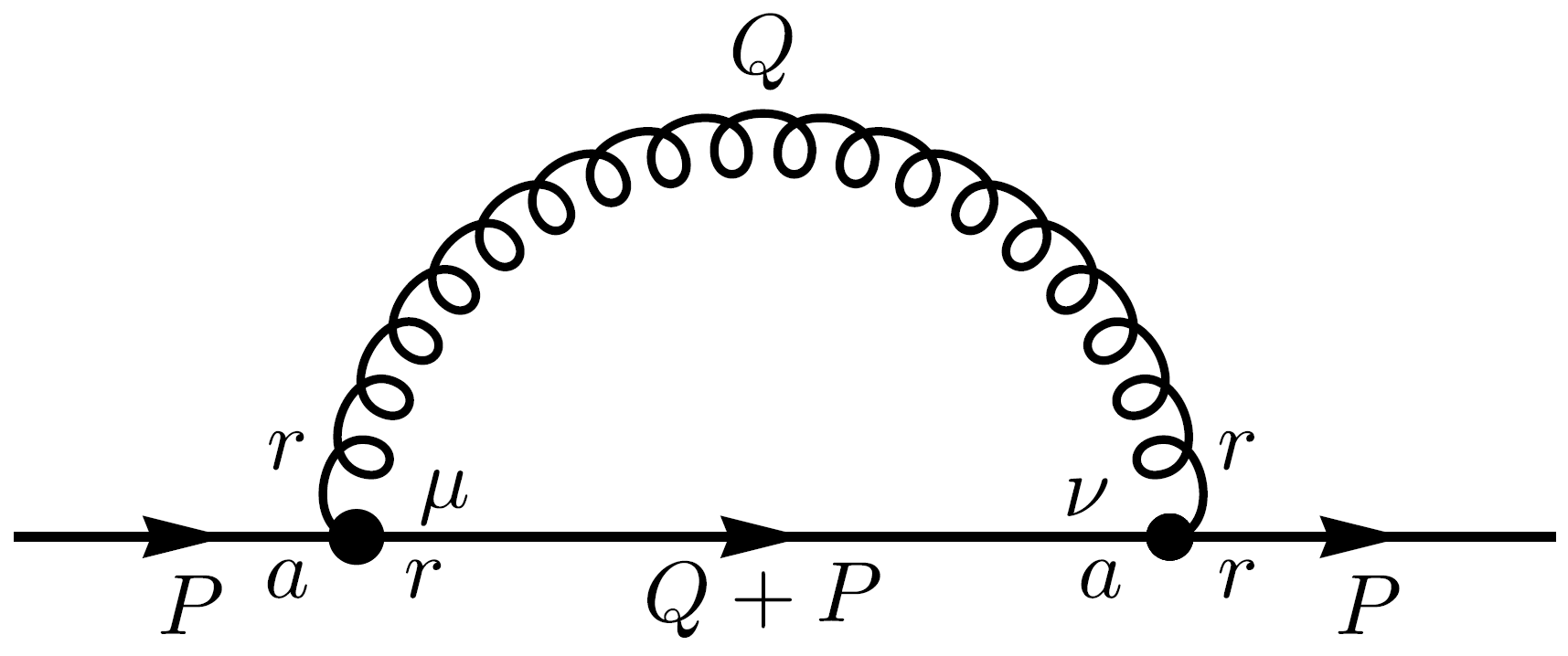}
    \includegraphics[width=0.4\linewidth]{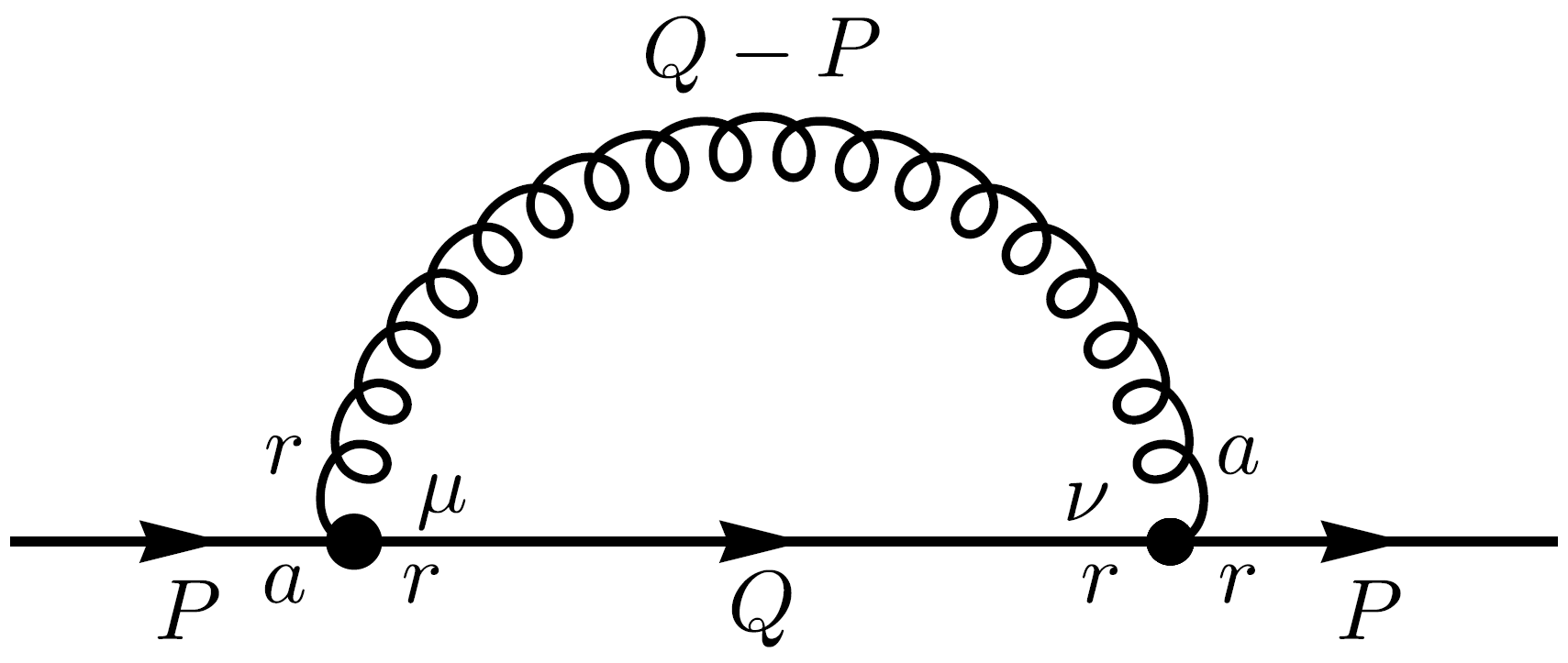}
    \caption{The retarded self-energies of fermions written in ra-basis describing equilibrium case. }
    \label{fig:self_energy}
\end{figure}

The retarded self-energies $\S_R$ and $\d\S_R$ are related to their counterpart in $ra$-basis as $\S_R=-i\S_{ar}$ and similarly for $\d\S_R$. We start with the evaluation of the equilibrium one. The diagrams in $ra$-basis are shown in Fig.~\ref{fig:self_energy}. The equilibrium quark and gluon propagators are given by
\begin{align}\label{Sra_explicit}
&S_{ra}(Q)=\frac{i{\slashed Q}}{(q_0+i\h)^2-\vec{q}^2},\nonumber\\
&S_{ar}(Q)=\frac{i{\slashed Q}}{(q_0-i\h)^2-\vec{q}^2},\nonumber\\
&S_{rr}(Q)=-2\p\e(q_0)\d(Q^2){\slashed Q}f(q_0),
\end{align}
and
\begin{align}\label{Dra_explicit}
&D_{\m\n}^{ra,AB}=\frac{i\d_{AB}}{(q_0+i\h)^2-\vec{q}^2}\(P_{\m\n}^T+\frac{Q^2u_\m u_\n}{q^2}\),\nonumber\\
&D_{\m\n}^{ar,AB}=\frac{i\d_{AB}}{(q_0-i\h)^2-\vec{q}^2}\(P_{\m\n}^T+\frac{Q^2u_\m u_\n}{q^2}\),\nonumber\\
&D_{\n\m}^{rr,AB}=2\p\d_{AB}\e(q_0)\d(Q^2)P_{\m\n}^T f(q_0)
\end{align}
respectively. We have used Coulomb gauge for the gluon propagator with $P_{\m\n}^T$ being the transverse projector, which contains spatial component only $P_{ij}^T=\d_{ij}-\hat{q}_i\hat{q}_j$. The vertices in $ra$-basis are easily derived easily as follows\footnote{We have absorbed $i^2$ into the definition of self-energy following \cite{Dong:2024cbt}}
\begin{align}\label{ra_vertices}
&g\bar{\ps}_1\g^\m\ps_1 t_AA_{\m,1}^A-g\bar{\ps}_2\g^\m\ps_2 t_AA_{\m,2}^A
\nonumber\\
&=g\bar{\ps}_r\g^\m\ps_r t_AA_{\m,a}^A+g\bar{\ps}_r\g^\m\ps_a t_AA_{\m,r}^A+g\bar{\ps}_a\g^\m\ps_r t_AA_{\m,r}^A-\frac{1}{4}g\bar{\ps}_a\g^\m\ps_a t_AA_{\m,a}^A.
\end{align}

We will consider an energetic quark with $P\gg T$ and $p_0>0$. It follows that the right diagram with on-shell quark in the loop is exponentially suppressed. We stress that the regime we consider is opposite to that of hard thermal loop (HTL), in which both diagrams contribute. The explicit representation of $\S_{ar}$ is given by
\begin{align}
\frac{\S_{ar}}{g^2C_F}&=\int_Q\g^\n S^{ra}(P-Q)\g^\m D^{rr}_{\n\m}(Q)\nonumber\\
=&\int_Q\g^\n\frac{i\g^\r(P-Q)_\r}{(P-Q)^2+i\e(p_0-q_0)\h}\g^\m\big[P_{\m\n}^T+\frac{Q^2}{q^2}u_\m u_\n\big]2\p\e(q_0)\d(Q^2)f(q_0)\nonumber\\
\simeq&\int_Q\g^\n\frac{i\g^\r P_\r}{P^2-2P\cdot Q+i\h}\g^\m P_{\m\n}^T 2\p\e(q_0)\d(Q^2)f(q_0),
\end{align}
where the Casimir comes from the color sum $C_F=t_At_A=\frac{N_c^2-1}{2N_c}$.
For on-shell gluon, we may write $P_{\m\n}^T=-\h_{\m\n}+\frac{Q_\m u_\n+Q_\n u_\m}{q_0}-\frac{Q_\m Q_\n}{q_0^2}$ and simplify the product of gamma matrices as
\begin{align}\label{gamma_product}
\g^\n{\slashed P}\g^\m P_{\m\n}^T
=2p_0\g^0-\frac{2}{q_0^2}q_j\g^i q_i q_j.
\end{align}
By rotational invariance, the relevant phase space integrals can be parameterized as
\begin{align}
&\int_Q\frac{q_i q_j}{P^2-2P\cdot Q+i\h}\frac{f(q_0)}{q_0^2}2\p\e(q_0)\d(Q^2)=A\d_{ij}+B\hat{p}_i\hat{p}_j,\nonumber\\
&\int_Q\frac{1}{P^2-2P\cdot Q+i\h}f(q_0)2\p\e(q_0)\d(Q^2)=C.
\end{align}
Contracting the first integral with $\d_{ij}$, we easily find $C=3A+B$. Thus
\begin{align}\label{Sigma_ar}
\frac{\S_{ar}}{g^2C_F}&=2ip_0\g^0(3A+B)-2ip_i\g^i(A+B)\nonumber\\
&=2i{\slashed P}(A+B)+4ip_0\g^0A.
\end{align}
The coefficients $A$ and $B$ are evaluated in appendix~\ref{sec_app_B} close to the mass shell with $P^2\ll p/\b$, for which we have
\begin{align}\label{coeff_ABC}
&3A+B=\frac{1}{2(2\p)^2}\frac{\p\(\p+2i\ln\frac{p\b(-1+a)}{2}\)}{4p\b},\nonumber\\
&A+B=\frac{1}{2(2\p)^2}\frac{\p\(\p+4ia+2i\ln\frac{p\b(-1+a)}{2}\)}{4p\b},
\end{align}
with $a=p_0/p+i\h$. The results are obtained by keeping the leading order terms in expansions in both $a-1$ and $(p\b)^{-1}$. It is instructive to compare our results with those in the HTL regime: firstly using \eqref{Sigma_ar} and \eqref{coeff_ABC}, we easily obtain the dispersion $p_0=p(1+8A)$ up to $O(g^2)$ with $A=\frac{1}{2(2\p)^2}\frac{-i\p}{2p\b}$. This indicates a finite damping rate in contrast to vanishing damping rate in the HTL case at $O(g^2)$; secondly $\S_R$ is complex for general $p_0$ in our case while the counterpart in HTL case is real except for spacelike momentum from Landau damping structures of branch cut also differ. The origin of the branch cut is also Landau damping, but the resulting branch cut lies at $a<1$ due to different kinematics. The branch cut appears when both particles in the loop go on shell, i.e. $(P-Q)^2=Q^2=0$, so that $P^2=2P\cdot Q$. Note that we have restricted ourselves to $(a-1)p\ll1/\b\sim q$. The condition reduces to $2P\cdot Q\simeq0$. It occurs at $a=\hat{q}\cdot\hat{p}$, which agrees with $a<1$.
\begin{figure}
    \centering
    \includegraphics[width=0.4\linewidth]{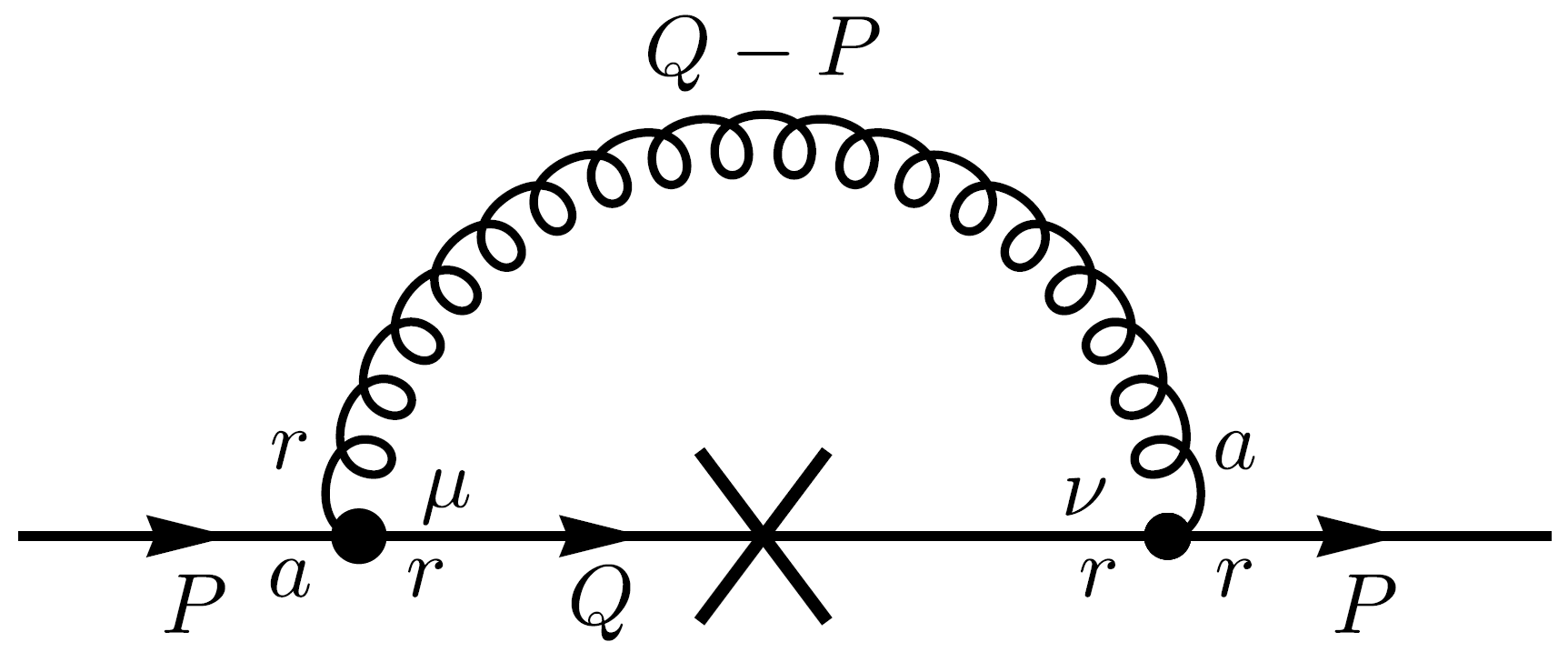}
    \includegraphics[width=0.4\linewidth]{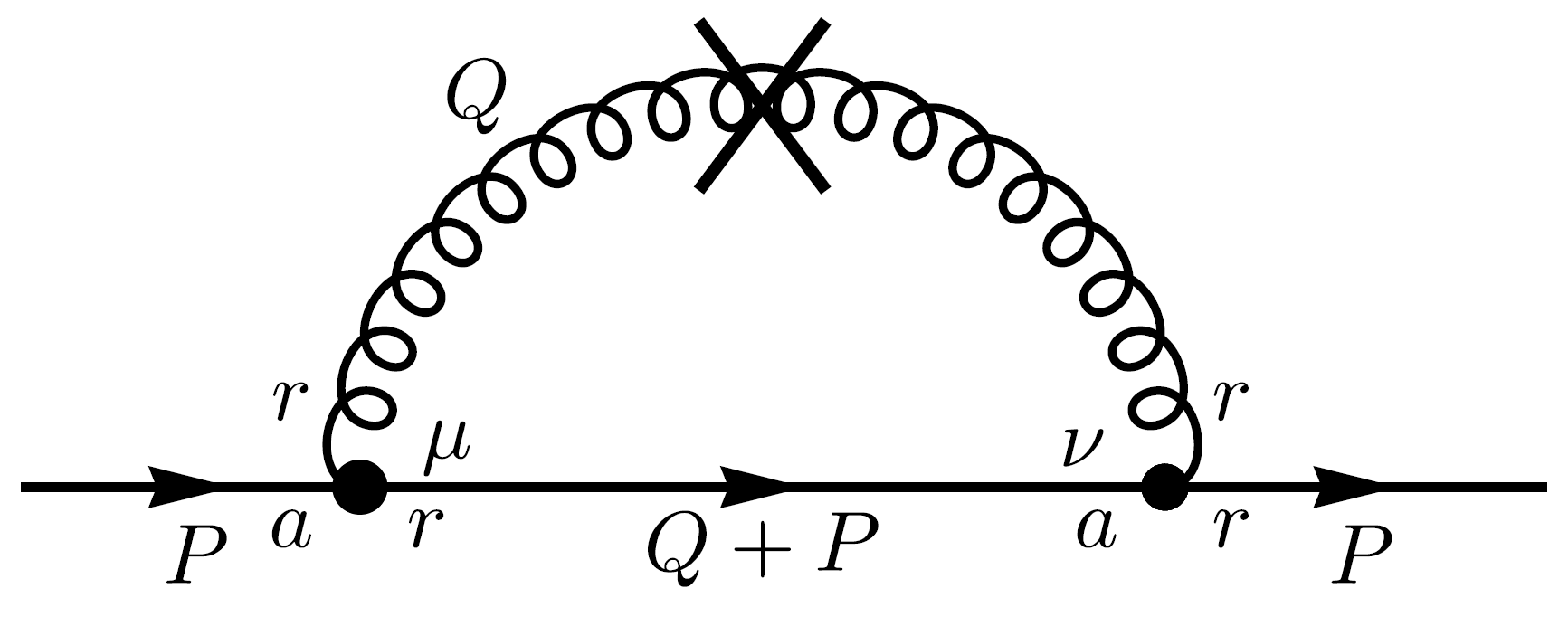}
    \caption{Self-energy of the fermions with the correction from off-equilibrium. The crosses in the diagrams denote the counterparts.}
    \label{fig:counterpart}
\end{figure}

We then turn to off-equilibrium correction to self-energy. At one-loop, the relevant diagrams are shown in Fig.~\ref{fig:counterpart}, which are the same as the diagrams in Fig.~\ref{fig:self_energy}, except with the equilibrium propagators replaced with off-equilibrium counterpart. 
Given that we aim at $\d\S_{ar}\sim O(g^2\pd)$ and the vertices contributing $g^2$ already, we only need to include $O(\pd)$ correction in the off-equilibrium propagators. These are given by the existing CKT framework. Note that in CKT, the spectral function is not corrected at $O(\pd)$ and by the KMS relation \eqref{spec_ra}, we deduce that $\d S_{ra}=0$. This can also be seen from \eqref{SR_sol} and \eqref{SR_sol2}. The same conclusion holds for gluons $\d D_{ra,AB}^{\m\n}=0$. The only $O(\pd)$ correction appears in $S_{rr}$ and $D_{rr,AB}^{\m\n}$. Below we relate $S_{rr}$ to the Wigner function known from CKT as an example.
The fields in the $ra$ basis are related to those in the Schwinger-Keldysh contour by
\begin{align}
\ps_r=\frac{1}{2}\(\ps_1+\ps_2\),\quad \ps_a=\ps_1-\ps_2,
\end{align}
and similarly for $\bar{\ps}$. We immediately have the following identities\footnote{In our convention, $S^<=-S_{12}$.}
\begin{align}\label{ra_12}
&S_{ra}=\frac{1}{2}\(S_{11}-S_{22}-S_{12}+S_{21}\),\nonumber\\
&S_{ar}=\frac{1}{2}\(S_{11}-S_{22}+S_{12}-S_{21}\),\nonumber\\
&S_{rr}=\frac{1}{4}\(S_{11}+S_{22}+S_{12}+S_{21}\),\nonumber\\
&S_{aa}=S_{11}+S_{22}-S_{12}-S_{21},
\end{align}
where the two-point function with indices $11$, $22$, $12$ and $21$ correspond to time-ordered, anti-time-ordered, lesser and greater functions respectively. One can easily verify the compatibility of \eqref{ra_def} with \eqref{ra_12}. The lesser function is just the Wigner function. The greater function is deduced as follows: $\r=S_{21}-S_{12}$. Again the spectral function is unmodified $\d\r=0$ at $O(\pd)$. We then find $\d S_{21}=\d S_{12}$. Furthermore, from the identity $S_{aa}=0$, we obtain $\d S_{rr}=\frac{1}{2}(\d S_{12}+\d S_{21})=\d S_{12}$.
We collect the explicit expressions for quark and gluon propagators from CKT \cite{Hidaka:2017auj,Huang:2020kik,Hattori:2020gqh}
\begin{align}\label{dSDra_explicit}
&\d S_{rr}(Q)=-2\p\e(q_0)\d(Q^2)\big[\frac{\e^{\m\n\a\b}Q_\a u_\b}{2q_0}\pd_\n\tilde{f}+Q^\m\frac{Q^\r\o_\r}{q_0}\b\frac{\pd}{\pd(\b q_0)}\tilde{f}\big]\g^5\g_\m,\nonumber\\
&\d D_{\n\m}^{rr,AB}=2\p\d_{AB}\e(q_0)\d(Q^2)\big[\frac{iP^T_{\n\a}Q^\a P_\m^{T,\b}}{2q_0^2}\pd_\b f-(\n\leftrightarrow\m)-i\frac{\e_{\n\m\a\b}Q^\a u^\b}{q_0^2}Q^\r\o_\r\b\frac{\pd}{\pd(\b q_0)}f\big].
\end{align}

We are ready to calculate the off-equilibrium correction to self-energy. For the same reason as its equilibrium counterpart, the diagram with on-shell quark in the loop is exponentially suppressed and not kept. The explicit representation of the other diagram in Fig.~\ref{fig:counterpart} is given by
\begin{align}\label{Sigma_rep}
&\frac{\d\S_{ar}}{g^2C_F}=\int_Q\g^\n S^{ar}(-P-Q)\g^\m \d D_{\n\m}^{rr}(Q)\nonumber\\
=&\int_Q\frac{\g^\n i\g^\r(P+Q)_\r\g^\m}{(P+Q)^2+i\e(p_0+q_0)\h}\big[\frac{iP^T_{\n\a}Q^\a P_\m^{T,\b}}{2q_0^2}\pd_\b f-(\n\leftrightarrow\m)-i\frac{\e_{\n\m\a\b}Q^\a u^\b}{q_0^2}Q^\r\o_\r\b f'\big]2\p\e(q_0)\d(Q^2)\nonumber\\
=&\int_Q\frac{\g^\n i\g^\r P_\r\g^\m}{P^2+2P\cdot Q+i\h}\big[\frac{iP^T_{\n\a}Q^\a P_\m^{T,\b}}{2q_0^2}\pd_\b f-(\n\leftrightarrow\m)-i\frac{\e_{\n\m\a\b}Q^\a u^\b}{q_0^2}Q^\r\o_\r\b f'\big]2\p\e(q_0)\d(Q^2).
\end{align}
We note that the square bracket is anti-symmetric in $\m\n$, so that there is only axial components in $\d\S_{ar}$ by using the identity
\begin{align}\label{gamma3}
\g^\m\g^\n\g^\r=\g^\m\h^{\n\r}-\g^\n\h^{\m\r}+\g^\r\h^{\m\n}-i\e^{\m\n\r\s}\g_5\g_\s.
\end{align}
Explicitly we have
\begin{align}
&\frac{\d\S_{ar}}{g^2C_F}=\int_Q\frac{\e^{ijk}\g^5\g_0 p_j}{P^2+2P\cdot Q+i\h}\big[\frac{iq_i}{q_0^2}(\b q_l\pd_k u_l+q_0\pd_k\b)f'\big]2\p\e(q_0)\d(Q^2)\nonumber\\
&-\int_Q\frac{2i\g^5\g_0(-p_l)}{P^2+2P\cdot Q+i\h}\frac{q_l}{q_0^2}(-q_m\o^m)\b f'2\p\e(q_0)\d(Q^2)\nonumber\\
&+\int_Q\frac{-\e^{ijk}\g^5\g_kp_0}{P^2+2P\cdot Q+i\h}\big[\frac{iq_i}{q_0^2}(\b q_l\pd_j u_l+q_0\pd_j\b)f'\big]2\p\e(q_0)\d(Q^2)\nonumber\\
&+\int_Q\frac{2\g^5\g_kp_0}{P^2+2P\cdot Q+i\h}\frac{iq_k}{q_0^2}(-q_m\o^m)\b f'2\p\e(q_0)\d(Q^2).
\end{align}
Following the same method as in the equilibrium case, we parametrize the relevant tensor integrals as
\begin{align}\label{tensors}
&\int_Q\frac{q_iq_j}{P^2+2P\cdot Q+i\h}\frac{f'(q_0)}{q_0^2}\e(q_0)\d(Q^2)=\d A\d_{ij}+\d B\hat{p}_i\hat{p}_j,\nonumber\\
&\int_Q\frac{q_i}{P^2+2P\cdot Q+i\h}\frac{f'(q_0)}{q_0}\e(q_0)\d(Q^2)=\d C\hat{p}_i.
\end{align}
In terms of the coefficients, we can express $\d\S_{ar}$ as
\begin{align}
&\frac{\d\S_{ar}}{g^2C_F}=2i\g^5\g_0\o^m p_m\b(-2\d A-\d B)+i\g^5\g_k\o^k\b p(-3 \d A)+i\e^{ijk}\g^5\g_k\hat{p}_i\hat{p}_l\pd_j u_l\b p(- \d B)\nonumber\\
&+i\e^{ijk}\g^5\g_k\hat{p}_i\pd_j\b p(- \d C)+i\g^5\g_k\hat{p}_k\hat{p}_m\o^m\b p(-2 \d B).
\end{align}
Using the relation $\pd_j u_l=\s_{jl}-\e^{jlk}\o^k$,\footnote{We have left out possible contribution from acceleration and $\pd\cdot u$. The former is not allowed in our approach and the latter does not couple to spin.} we arrive at
\begin{align}\label{Sigma_final}
&\frac{\d\S_{ar}}{g^2C_F}=2i\g^5\g_0\o^m p_m\b(-2\d A -\d B)+i\g^5\g_k\o^k\b p(-4\d A- \d B)+i\e^{ijk}\g^5\g_k\hat{p}_i\hat{p}_l\s_{jl}\b p(- \d B)\nonumber\\
&+i\e^{ijk}\g^5\g_k p_i\pd_j\b (-\d C)+i\g^5\g_k\hat{p}_k p_m\o^m\b(-\d B)\nonumber\\
&=2i\g^5\g_0\o^m p_m\b (-2\d A -\d B )+i\g^5\g_k\o_\parallel^k\b p(-4 \d A-2\d B)+i\g^5\g_k\o_\perp^k\b p(-4\d A-\d B)\nonumber\\
&+i\e^{ijk}\g^5\g_k\hat{p}_i\hat{p}_l\s_{jl}\b p(-\d B)
+i\e^{ijk}\g^5\g_k p_i\pd_j\b (-\d C).
\end{align}
In the last step, we have defined $\o_\parallel^k=\o^i\hat{p}_i\hat{p}^k$, $\o_\perp^k=\o^k-\o_\parallel^k$. $\d\S_{ar}$ contains axial components only and can be decomposed as
\begin{align}\label{axial}
&\d\S_{ar}=\g^5\g^\m{\cal A}_\m,\nonumber\\
&{\cal A}^0=i\o^i p_i\b(-4\d A-2\d B),\nonumber\\
&{\cal A}^k=i\o^k_\parallel\b p(-4\d A-2\d B)+\o_\perp^k\b p(-4\d A-\d B)+\e^{ijk}\hat{p}_i\hat{p}_l\s_{jl}\b p(-\d B)+\e^{ijk}p_i\pd_j\b (-\d C).
\end{align}
A notable property is that $P_\m {\cal A}^\m=0$. 
The coefficients $A$, $B$ and $C$ have been evaluated in appendix \ref{sec_app_B} with the following results to leading order in $T/p$:
\begin{align}\label{coeff_dABC}
&3\d A+\d B=\frac{1}{4(2\p)^2}\(-\frac{2i\p}{(a^2-1)p^2\b^2}+\frac{2C_b+i\p C_a-2C_a\ln\frac{p\b(-1+a)}{2}}{2p\b}\),\nonumber\\
&\d A+\d B=\frac{1}{4(2\p)^2}\(-\frac{2i\p}{3(a^2-1)p^2\b^2}+\frac{-4C_a+2C_b+i\p C_a-2C_a\ln\frac{p\b(-1+a)}{2}}{2p\b}\),\nonumber\\
&\d C=\frac{1}{4(2\p)^2}\(\frac{-4C_a+2C_b+i\p C_a-2C_a\ln\frac{p\b(-1+a)}{2}}{2p\b}\).
\end{align}

Now we are ready to plug \eqref{Sigma_ar} and \eqref{axial} into \eqref{SR_sol2}. We first look at the first term of \eqref{SR_sol2}, which can be simplified using $P\cdot{\cal A}=0$ as
\begin{align}
\d S_R^\bzero=-\frac{1}{{\slashed P}}\d\S_R\frac{1}{{\slashed P}}=\frac{1}{P^2}\d\S_R,
\end{align}
To compare with known contribution in free theory, which is localized at $p_0=p$, we integrate over $p_0$. Since the explicit results \eqref{coeff_dABC} have been obtained assuming $(a-1)p\ll 1/\b$, we integrate over $-\L+p,\L+p$ with $\L\lesssim 1/\b$ to obtain the following integrals needed for our purpose
\begin{align}\label{p0_integrals}
&\int_{p-\L}^{p+\L} dp_0\frac{1}{P^2}\ln\frac{p_0-p+i\h}{p}\simeq\int_{p-\L}^{p+\L} dp_0\frac{1}{2p}\frac{1}{p_0-p+i\h}\ln\frac{p_0-p+i\h}{p}=\frac{1}{2p}\(\frac{\p^2}{2}+i\p\ln (p\b)\),\nonumber\\
&\int_{p-\L}^{p+\L} dp_0\frac{1}{P^2}\simeq\int_{p-\L}^{p+\L} dp_0\frac{1}{2p}\frac{1}{p_0-p+i\h}=\frac{1}{2p}(-i\p),\nonumber\\
&\int_{p-\L}^{p+\L} dp_0\frac{i}{(P^2)^2}\simeq\int_{p-\L}^{p+\L} dp_0\frac{1}{(2p)^2}\frac{i}{(p_0-p+i\h)^2}=\frac{1}{(2p)^2}\(\frac{2}{\L}\).
\end{align}
It is worth noting that the first two integrals are independent of $\L$, which indicates a localized contribution. On the contrary the third integral does depend on $\L$. Interestingly, the integrand of the latter is actually exact (see discussion below \eqref{below_qst_off}). This allows us to set $\L=\# p\gg 1/\b$,\footnote{The constant $\#$ cannot be large in order not to pick up contribution from the anti-particle.} leading to a result that is suppressed by $p\b$ compared to other contributions. We drop the corresponding contribution below. Using $\d\S_R=-i\d\S_{ar}$ and \eqref{p0_integrals}, we find terms logarithmic in $p\b$ cancel and finally arrive at
\begin{align}\label{SR0_final}
&\int dp_0\d S_R^\bzero=\g^5\g^\m{\cal B}_\m,\nonumber\\
&\text{Im}[{\cal B}^0]=\frac{g^2C_F}{4(2\p)^2}\o^i\hat{p}_i\frac{\p}{2p}(2C_b+(-2+\ln 4)C_a),\nonumber\\
&\text{Im}[{\cal B}^k]=\frac{g^2C_F}{4(2\p)^2}\frac{\p}{2p}\big[\o_\parallel^k(2C_b+(-2+\ln 4)C_a)+\frac{2C_b+(2+\ln 4)C_a}{2}\o_\perp^k+\nonumber\\
&\frac{2C_b+(-6+\ln 4)C_a}{2}\e^{ijk}\hat{p}_i\hat{p}_l\s_{jl}+\frac{2C_b+(-4+\ln 4)C_a}{2}\e^{ijk}\hat{p}_i\frac{\pd_j\b}{\b}\big].
\end{align}

We then turn to the second term of \eqref{SR_sol2}, which is written explicitly as
\begin{align}\label{SR1}
S_R^\bone=\g^5\g_\b P_\n\pd_\m \S^R_\l\frac{-1}{(P^2)^2}\e^{\b\l\m\n},
\end{align}
with
\begin{align}\label{Sigma_lambda}
\S_R^\l=(\S_R^0,\S_R^i)=\S_R^i\hat{p}_i\frac{1}{p}P^\l+\D\S u^\l,
\end{align}
with $\D\S=\S_R^0-\S_R^i\hat{p}_i\frac{p_0}{p}$.
Note that the first term in \eqref{Sigma_lambda} does not contribute to \eqref{SR1}, and the second term gives
\begin{align}
S_R^\bone=\g^5\g_\b P_\n\pd_\m (\D\S u_\l)\frac{-1}{(P^2)^2}\e^{\b\l\m\n}.
\end{align}
The coordinate dependence of $\D\S$ comes from local fluid velocity and temperature. The former determines components of momentum as $p_0=P\cdot u$ and $p=(p_0^2-P^2)^{1/2}$. Thus
\begin{align}
\pd_\m\D\S&=\frac{\pd\D\S}{\pd p_0}\pd_\m p_0+\frac{\pd\D\S}{\pd p}\pd_\m p+\frac{\pd\D\S}{\pd\b}\pd_\m\b\nonumber\\
&=\(\frac{\pd\D\S}{\pd p_0}+\frac{\pd\D\S}{\pd p}\frac{p_0}{p}\)\pd_\m p_0-\D\S\frac{\pd_\m\b}{\b}\nonumber\\
&=\D\S^\pd_p P^\a\pd_\m u_\a-\D\S\frac{\pd_\m\b}{\b},
\end{align}
where in the second line we have used the fact that $\D\S\propto\b^{-1}$ based on \eqref{coeff_ABC}. Specializing to static off-equilibrium state, we can write out $S_R^\bone$ explicitly as
\begin{align}\label{SR1_eq}
S_R^\bone&=\big[2\g^5(\g_0p_i+\g_i p_0)\D\S\o^i+\g^5\g_i\D\S^\pd_p\(p_lp_j\s_{kl}\e^{ijk}-p_ip_m\o^m+p^2\o^i\)-\g^5\g_i p_j\frac{\pd_k\b}{\b}\D\S\e^{ijk}\big]\nonumber\\
&\times\frac{1}{(P^2)^2}.
\end{align}
The explicit expressions for $\D\S$ and $\D\S^\pd_p$ are worked out as
\begin{align}
&\D\S=\frac{2a}{(2\p)^2\b}\(-\frac{i\p}{2}\),\nonumber\\
&\D\S^\pd_p=\frac{1}{p}\frac{\pd\D\S}{\pd a}(1-a^2)=\frac{2(1-a^2)}{(2\p)^2p\b}\(-\frac{i\p}{2}\).
\end{align}
An important property is that both $\D\S$ and $\D\S^\pd_p$ are regular as $a\to 1$, so that here we can simply use $a=p_0/p$ to proceed. The following integrals are needed in the evaluation of $S_R^{(1)}$
\begin{align}
&\int_{p-\L}^{p+\L}dp_0\frac{a^2-1}{(P^2)^2}\simeq\frac{1}{p^2}\int_{p-\L}^{p+\L}dp_0\frac{1}{P^2}=\frac{1}{p^2}\frac{1}{2p}(-i\p),\nonumber\\
&\int_{p-\L}^{p+\L}dp_0\frac{a}{(P^2)^2}\simeq\frac{1}{p}\int_{-\L}^{\L}dp_0\frac{p_0}{(P^2)^2}\sim \frac{1}{p^2\L},
\end{align}
In the last line we have again set $\L=\# p$ because $\D\S$ is exact like the off-equilibrium counterpart. Using \eqref{SR1_eq} and the integrals above, we find the corresponding contribution to $\int dp_0 S_R^\bone$ is parametrically suppressed compared to \eqref{SR0_final}. The reason can be traced in different temperature dependence of the equilibrium and off-equilibrium self-energies: the former is linear in temperature while the latter is dominated by temperature independent contribution in the regime $p\b\gg1$ we choose to work. It follows the equilibrium contribution to spectral function is suppressed.

In summary, the radiative correction to spectral function is given by
\begin{align}
&\int dp_0\d\r(P)=2\text{Im}\[\d S_R^\bzero+S_R^\bone\]\nonumber\\
&=\frac{g^2C_F}{2(2\p)^2}\frac{\p}{2p}\g^5\g_i\big[\o_\parallel^i(2C_b+(-2+\ln 4)C_a)+\frac{2C_b+(2+\ln 4)C_a}{2}\o_\perp^i+\nonumber\\
&\frac{2C_b+(-6+\ln 4)C_a}{2}\e^{ijk}\hat{p}_j\hat{p}_l\s_{kl}+\frac{2C_b+(-4+\ln 4)C_a}{2}\e^{ijk}\hat{p}_j\frac{\pd_k\b}{\b}\big].
\end{align}
The fact that the spectral function is modified by hydrodynamic gradient indicates that the particle itself is polarized.
Using the equilibrium KMS relation, we obtain the following contribution to the spin polarization
\begin{align}\label{pol_radiative}
&\int dp_0\d S^<(P)=\int dp_0 \d\r(P) f(p_0)\nonumber\\
&=\frac{g^2C_F}{2(2\p)^2}\frac{\p}{2p}\g^5\g_i\big[0.95\o_\parallel^i+1.48\o_\perp^i-0.52\e^{ijk}\hat{p}_j\hat{p}_l\s_{kl}-0.02\e^{ijk}\hat{p}_j\frac{\pd_k\b}{\b}\big]f(p),
\end{align}
where we have used $C_a=0.5$ and $C_b\simeq 0.63$ from appendix~\ref{sec_app_B}.
This is to be compared with the tree-level contribution from the same sources \eqref{Wigner_rep}
\begin{align}\label{pol_tree}
\int dp_0 S_{(0)}^<(P)&=-\g^5\g_i\frac{2\p}{2}\(\o^i+\e^{ijk}\hat{p}_k\frac{\pd_j\b}{\b}+\e^{ijk}\hat{p}_k\hat{p}_l\s_{lj}\)\frac{\pd}{\pd(\b p)}f(p)\nonumber\\
&\simeq \g^5\g_i\frac{2\p}{2}\(\o^i+\e^{ijk}\hat{p}_k\frac{\pd_j\b}{\b}+\e^{ijk}\hat{p}_k\hat{p}_l\s_{lj}\)f(p),
\end{align}
where we have used $\frac{\pd}{\pd(\b p)}f(p)\simeq -f(p)$ in the regime $p\b\gg1$.

\section{Outlook}\label{sec_outlook}

We have scrutinized the idea of mimicking off-equilibrium state effect by metric perturbation on equilibrium state. We have identified two effects of metric perturbation: one is to scatter particle in the background metric, which we find reproduces the affine connection contribution in CKT and also captures spin response to vorticity. Together with gauge link contribution, it is possible to describe spin response to general hydrodynamic gradient. However the description is limited by kinematic condition in the scattering picture, which cannot distinguish vorticity and shear. The other is genuine off-equilibrium effect. We find a suitable choice of equilibrium state in curved space allows us to mimic off-equilibrium state in flat space. The latter allows us to study spin responses to vorticity, shear and temperature gradient unambiguously.

We have classified general off-equilibrium contribution into three types including modification to (i) spectral function; (ii) distribution function; (iii) KMS relation. We have then considered radiative correction to type (i) and found the correction leads to a polarized spectral function in an off-equilibrium medium. Clearly it would be desirable to study radiative correction type (ii) and (iii). For correction of type (ii), we know particle scattering on metric field can capture spin-vorticity. Therefore, we can study the corresponding radiative correction using in-medium gravitational form factors \cite{Lin:2023ass}. Correction of type (iii) would involve gradient correction to Kadanoff-Baym equation. We leave these interesting directions for future exploration.

\section*{Acknowledgments}
We are grateful to X.-G. Huang, K. Mameda and Y. Yin for stimulating discussions. This work is in part supported by NSFC under Grant Nos 12075328, 11735007.

\appendix

\section{Solving the Kadanoff-Baym equation}\label{sec_app_A}

In this appendix, we solve \eqref{KB} order by order in gradient
\begin{align}
S_R=S_R^\bzero+S_R^\bone+\cdots,
\end{align}
with superscript denoting the order. Each order solution is further expanded in self-energy. $S_R^\bzero$ is solved as
\begin{align}\label{S_R0}
{\slashed P}S_R^\bzero-\S_R S_R^\bzero=-1\Rightarrow S_R^\bzero=-\frac{1}{{\slashed P}}-\frac{1}{{\slashed P}}\S_R\frac{1}{{\slashed P}}+\cdots,
\end{align}
$S_R^\bone$ satisfies the following equation
\begin{align}\label{S_R1_eq}
{\slashed P}S_R^\bone-\S_R S_R^\bone=-\frac{i}{2}{\slashed\pd}S_R^\bzero+\frac{i}{2}\{\S_R,S_R^\bzero\}.
\end{align}
We shall find solution of $S_R^\bone$ only to lowest order in $\S_R$, which is counted as $O(g^2)$. To this end, we may drop the self-energy term on the LHS and keep terms up to $O(g^2\pd)$ on the RHS. We simplify the RHS as
\begin{align}
&\frac{i}{2}{\slashed\pd}\(\frac{1}{{\slashed P}}\S_R\frac{1}{{\slashed P}}\)-\frac{i}{2}\pd\S_R\cdot\pd_P\(\frac{-1}{{\slashed P}}\)\nonumber\\
=&\frac{i}{2}{\slashed\pd}\S_R\frac{1}{P^2}-\frac{i}{2}{\slashed\pd}\big[\frac{1}{{\slashed P}},\S_R\big]\frac{1}{{\slashed P}}+\frac{i}{2}\S_R\overleftarrow{\pd}-i\pd_\l\S_R{\slashed P}\frac{P^\l}{(P^2)^2}\nonumber\\
=&i\pd\cdot\S_R\frac{1}{P^2}+i{\slashed\pd}\S^R_\l\({\slashed P}P^\l-\g^\l P^2\)\frac{1}{(P^2)^2}-i\pd_\l\S_R{\slashed P}\frac{P^\l}{(P^2)^2}\nonumber\\
=&-\s^{\m\l}\pd_\m\S^R_\l\frac{1}{P^2}+i\g^\m\(\pd_\m\S^R_\l\-(\l\leftrightarrow\m)\){\slashed P}\frac{P^\l}{(P^2)^2}\nonumber\\
=&-\s^{\m\n}T_{\m\n}\frac{1}{P^2}+2\s^{\m\n}P_\n T_{\m\l}\frac{P^\l}{(P^2)^2},
\end{align}
where we have used the product $\g^\m\g^\n=\h^{\m\n}-i\s^{\m\n}$ and defined $T_{\m\l}=\pd_{[\m}\S^R_{\l]}$.

Motivated by solution in CKT without radiative correction, we solve \eqref{S_R1_eq} with the following ansatz $S_R^\bone=\g^5\g^\n{\cal A}_\n$. Using the following relation
\begin{align}
{\slashed P}\g^5\g^\n{\cal A}_\n=-\g^5P\cdot {\cal A}+i\g^5\s^{\m\n}P_\m{\cal A}_\n=-\g^5P\cdot {\cal A}-\frac{1}{2}\e^{\m\n\a\b}\s_{\a\b}P_\m{\cal A}_\n.
\end{align}
We proceed by dropping the $P\cdot {\cal A}$ term in the above, whose validity will be verified below. Extracting coefficients of $\s^{\m\n}$ from \eqref{S_R1_eq}, we obtain
\begin{align}
&P_\a{\cal A}_\b-(\a\leftrightarrow\b)=\(-T^{\m\n}\frac{1}{P^2}+2P^\n T^{\m\l}\frac{P_\l}{(P^2)^2}\)\e_{\m\n\a\b}\nonumber\\
=&-T^{\m\n}\frac{1}{P^2}\e_{\m\n\a\b}+2P^\n T^{\m\l}\frac{-1}{(P^2)^2}\(P_\n\e_{\a\b\l\m}+P_\a\e_{\b\l\m\n}+P_\b\e_{\l\m\n\a}\)\nonumber\\
=&P^\n T^{\m\l}\frac{-1}{(P^2)^2}\(P_\a\e_{\b\l\m\n}+P_\b\e_{\l\m\n\a}\).
\end{align}
It is solved by ${\cal A}_\b=P^\n T^{\m\l}\frac{-1}{(P^2)^2}\e_{\b\l\m\n}$. It indeed satisfies $P\cdot{\cal A}=0$, justifying dropping the corresponding term above.

\section{Evaluation of momentum integrals}\label{sec_app_B}
We first consider the following tensor integral for equilibrium self-energy
\begin{align}
\int_Q\frac{q_i q_j}{P^2-2P\cdot Q+i\h}\frac{f(\b q_0)}{q_0^2}2\p\e(q_0)\d(Q^2)=A\d_{ij}+B\hat{p}_i\hat{p}_j.
\end{align}
Contracting the above with $\d^{ij}$ and $\hat{p}_i\hat{p}_j$ respectively, we find
\begin{align}
&3A+B=\int_Q\frac{q^2}{P^2-2P\cdot Q+i\h}\frac{f(\b q_0)}{q_0^2}2\p\e(q_0)\d(Q^2),\nonumber\\
&A+B=\int_Q\frac{(q\cdot\hat{p})^2}{P^2-2P\cdot Q+i\h}\frac{f(\b q_0)}{q_0^2}2\p\e(q_0)\d(Q^2).
\end{align}
We shall illustrate the evaluation of the first integral. The same method applies to the second integral as well. We first perform the integration of $q_0$ using $\d(Q^2)$ to obtain
\begin{align}
\int\frac{d^3q}{(2\p)^3}\(\frac{1}{P^2-2(p_0q-\vec{p}\cdot\vec{q})+i\h}\frac{f(\b q)}{2q}-\frac{1}{P^2-2(-p_0q-\vec{p}\cdot\vec{q})+i\h}\frac{f(-\b q)}{2q}\).
\end{align}
We use $f(-\b q)=-1-f(\b q)$ and note that the $-1$ gives rise to vacuum self-energy, which will be eliminated by spacetime derivatives, so that we can drop the $-1$. We then perform angular integration to obtain
\begin{align}\label{angular}
\int\frac{dq}{2(2\p)^2}\frac{1}{2pq}\(\ln\frac{p^2(a^2-1)-2pq(a-1)}{p^2(a^2-1)-2pq(a+1)}+\ln\frac{p^2(a^2-1)+2pq(a+1)}{p^2(a^2-1)+2pq(a-1)}\)\frac{f(\b q)}{q},
\end{align}
with $a=p_0/p$ and $i\h$ replaced by an infinitesimal positive imaginary part of $a$.
The $q$-integration cannot be performed analytically. To proceed, we restrict ourselves to the regime $(a-1)p\ll 1/\b\ll p$, or equivalently and $P^2\ll p/\b\ll p^2$, that is an energetic particle close to the mass shell. This allows us to split the integration domain into two parts separated by a floating point $\qst$, with $(a-1)p\ll\qst\ll 1/\b$. For $0<q<\qst$, we use $f(q)\simeq\frac{1}{\b q}$ to have
\begin{align}\label{below_qst}
&\int_0^{\qst} dq\frac{1}{2pq}\(\ln\frac{P^2-2pq(a-1)+i\h}{P^2-2(a+1)pq+i\h}+\ln\frac{P^2+2pq(a+1)+i\h}{P^2+2(a-1)pq+i\h}\)\frac{q^2}{-q}\frac{1}{\b q}\nonumber\\
\simeq&\frac{1}{4p\b}\(\p^2+2i\p\ln\frac{(-1+a)p}{2\qst}\).
\end{align}
The LHS can be integrated analytically with a lengthy expression not shown here. The RHS is obtained by keeping the leading term in the expansion in $a(p-1)/\qst$ and then keeping the leading term in the expansion in $\qst/p$. For $\qst<q<\infty$, we apply the same approximation to the integrand of \eqref{angular} excluding the $f(\b q)$ factor to obtain
\begin{align}
&\frac{q^2}{2pq}\(\ln\frac{p^2(a^2-1)-2pq(a-1)}{p^2(a^2-1)-2pq(a+1)}+\ln\frac{p^2(a^2-1)+2pq(a+1)}{p^2(a^2-1)+2pq(a-1)}\)\frac{1}{2q}\nonumber\\
\simeq&\frac{\ln(1-a)-\ln(-1+a)}{2p}=-\frac{i\p}{2p}.
\end{align}
An infinitesimal positive imaginary part of $a$ has been invoked in the last equality. We then convolute the above with $f(q)$ and expanding in $\b\qst$ to obtain
\begin{align}\label{above_qst}
-\frac{i\p}{2p}\int_{\qst}^\infty f(\b q)\simeq\frac{i\p}{2p}\frac{\ln(\qst\b)}{\b}.
\end{align}
Adding up \eqref{below_qst} and \eqref{above_qst}, we find the dependencies on the floating scale $\qst$ cancel to give
\begin{align}\label{3AplusB}
3A+B=\frac{1}{2(2\p)^2}\frac{\p\(\p+2i\ln\frac{p\b(-1+a)}{2}\)}{4p\b}.
\end{align}
Repeating the same procedure as above, we obtain for the second integral
\begin{align}\label{AplusB}
A+B=\frac{1}{2(2\p)^2}\frac{\p\(\p+4i+2i\ln\frac{p\b(-1+a)}{2}\)}{4p\b}.
\end{align}

Now we turn to integrals for off-equilibrium self-energy. Contracting the first tensor integral of \eqref{tensors} by $\d^{ij}$ and $\hat{p}_i\hat{p}_j$ and contracting the second vector integral by $\hat{p}_i$, we obtain
\begin{align}
&3\d A+\d B=\int_Q\frac{q^2}{P^2+2P\cdot Q_i\h}\frac{f'(\b q_0)}{2q_0^2}\e(q_0)\d(Q^2),\nonumber\\
&\d A+\d B=\int_Q\frac{(q\dot\hat{p})^2}{P^2+2P\cdot Q_i\h}\frac{f'(\b q_0)}{2q_0^2}\e(q_0)\d(Q^2),\nonumber\\
&\d C=\int_Q\frac{q\dot\hat{p}}{P^2+2P\cdot Q_i\h}\frac{f'(\b q_0)}{2q_0^2}\e(q_0)\d(Q^2).
\end{align}
We shall illustrate the evaluation of the first integral. As before, we integrate over $q_0$ using the Dirac delta function and then integrate over the angular variables to obtain
\begin{align}\label{angular_off}
\frac{1}{2pq}\(-\ln\frac{p^2(a^2-1)-2pq(a-1)}{p^2(a^2-1)-2pq(a+1)}+\ln\frac{p^2(a^2-1)+2pq(a+1)}{p^2(a^2-1)+2pq(a-1)}\)q f'(q).
\end{align}
We have used $f'(-\b q)=f'(\b q)$. Note that there is no vacuum contribution in off-equilibrium self-energy. Following the equilibrium case, we split the integration domain into two parts with a floating scale $\qst$. The part below $\qst$ is obtained by the approximation $f'(\b q)\simeq -\frac{1}{(\b q)^2}$ as
\begin{align}\label{below_qst_off}
&\int_0^{\qst}dq \frac{1}{2pq}\(-\ln\frac{p^2(a^2-1)-2pq(a-1)}{p^2(a^2-1)-2pq(a+1)}+\ln\frac{p^2(a^2-1)+2pq(a+1)}{p^2(a^2-1)+2pq(a-1)}\)q\(-\frac{1}{(\b q)^2}\)\nonumber\\
=&F(\qst)-F(0)\nonumber\\
\simeq&-\frac{2i\p}{(a^2-1)p^2\b^2}+\frac{2+i\p-2\ln\frac{p(-1+a)}{2\qst}}{2p\qst\b^2},
\end{align}
where $F$ is the integrated function. $F(0)=\frac{2i\p}{(a^2-1)p^2\b^2}$ is exact. $F(\qst)$ is truncated to leading order expansions in $(a-1)$ and $\qst/p$.
The part above $\qst$ is evaluated using the following approximation for the integrand excluding the $f'(\b q)$ factor
\begin{align}
&\frac{1}{2pq}\(-\ln\frac{p^2(a^2-1)-2pq(a-1)}{p^2(a^2-1)-2pq(a+1)}+\ln\frac{p^2(a^2-1)+2pq(a+1)}{p^2(a^2-1)+2pq(a-1)}\)q\nonumber\\
\simeq&\frac{i\p+2\ln\frac{2q}{p(-1+a)}}{2p}.
\end{align}
The convolution with $f'(\b q)$ is performed as follows
\begin{align}\label{above_qst_off}
&\int_{\qst}^\infty dq \frac{i\p+2\ln\frac{2q}{p(-1+a)}}{2p}f'(\b q)\nonumber\\
=&\frac{i\p+2\ln\frac{2}{p\b(-1+a)}}{2p\b}\int_{x_0}^\infty dx f'(x)+\frac{1}{p\b}\int_{x_0}^\infty f'(x)\ln x,
\end{align}
where $x=q\b$ and $x_0=\qst\b$. Since $x_0\ll 1$, the $x$-integrals can be expanded as
\begin{align}\label{x-integrals}
&\int_{x_0}^\infty dx f'(x)=-\frac{1}{x_0}+C_a,\nonumber\\
&\int_{x_0}^\infty dx f'(x)\ln x=-\frac{1}{x_0}(1+\ln x_0)+C_b,
\end{align}
with $C_a=0.5$ and $C_b\simeq 0.630$, which are calculated numerically. Combining \eqref{below_qst_off} and \eqref{above_qst_off}, we again find the expected cancellation on the dependencies on $\qst$:
\begin{align}\label{3dAplusdB}
3\d A+\d B=\frac{1}{4(2\p)^2}\(-\frac{2i\p}{(a^2-1)p^2\b^2}+\frac{2C_b+i\p C_a-2C_a\ln\frac{p\b(-1+a)}{2}}{2p\b}\).
\end{align}
The other two integrals can be calculated similarly to give
\begin{align}\label{dAplusdB_dC}
&\d A+\d B=\frac{1}{4(2\p)^2}\(-\frac{2i\p}{3(a^2-1)p^2\b^2}+\frac{-4C_a+2C_b+i\p C_a-2C_a\ln\frac{p\b(-1+a)}{2}}{2p\b}\),\nonumber\\
&\d C=\frac{1}{4(2\p)^2}\(\frac{-4C_a+2C_b+i\p C_a-2C_a\ln\frac{p\b(-1+a)}{2}}{2p\b}\).
\end{align}



\bibliographystyle{unsrt}\bibliography{references.bib}

\end{CJK}

\end{document}